\documentclass[aps,superscriptaddress,twocolumn,a4paper,10pt]{revtex4}
\usepackage[colorlinks=true, pdfstartview=FitV, linkcolor=blue, citecolor=red, urlcolor=black]{hyperref}
\usepackage{graphicx}
\usepackage[all]{xy}
\usepackage{amsmath}
\usepackage{amssymb}
\usepackage{tensor}
\usepackage{orcidlink}
\usepackage{amsthm}
\usepackage{comment}
\newcommand{\be}{\begin{equation}}
\newcommand{\ee}{\end{equation}}
\newcommand{\ben}{\begin{eqnarray}}
\newcommand{\een}{\end{eqnarray}}
\newcommand{\bes}{\begin{subequations}}
\newcommand{\ees}{\end{subequations}}
\def\bal#1\eal{\begin{align}#1\end{align}}

\newcommand{\bfi}{\begin{figure}}
\newcommand{\efi}{\end{figure}}
\newcommand{\bc}{\begin{center}}
\newcommand{\ec}{\end{center}}

\newcommand{\vphi}{{\varphi}}
\newcommand{\sech}{{\rm sech}}

\newcommand{\arcsinh}{{\rm arcsinh}}
\newcommand{\p}{{\partial}}

\begin{document}

\title{Super long-range kinks}
%-------------------------------------------%
\author{I. Andrade\,\orcidlink{0000-0002-9790-684X}}
        \email[]{andradesigor0@gmail.com}\affiliation{Departamento de F\'\i sica, Universidade Federal da Para\'\i ba, 58051-970 Jo\~ao Pessoa, PB, Brazil}
\author{M.A. Marques\,\orcidlink{0000-0001-7022-5502}}
        \email[]{marques@cbiotec.ufpb.br}\affiliation{Departamento de Biotecnologia, Universidade Federal da Para\'\i ba, 58051-900 Jo\~ao Pessoa, PB, Brazil}
\author{R. Menezes\,\orcidlink{0000-0002-9586-4308}}
     \email[]{rmenezes@dcx.ufpb.br}\affiliation{Departamento de Ci\^encias Exatas, Universidade Federal
da Para\'{\i}ba, 58297-000 Rio Tinto, PB, Brazil}\affiliation{Departamento de F\'{\i}sica, Universidade Federal de Campina Grande,  58109-970 Campina Grande, PB, Brazil}

\begin{abstract}
In this work we investigate the presence of scalar field models supporting kink solutions with logarithmic tails, which we call super long-range structures. We first consider models with a single real scalar field and associate the long-range profile to the orders of vanishing derivatives of the potential at its minima. We then present a model whose derivatives are null in all orders and obtain analytical solutions with logarithmic falloff. We also show that these solutions are stable under small fluctuations. To investigate the forces between super long-range structures, we consider three methods and compare them. Next, we study two-field models in which the additional field is used to modify the kinetic term of the other. By using a first-order formalism based on the minimization of the energy, we explore the situation in which one of the fields can be obtained independently from the other. Within this framework, we unveil how to smoothly go from long- or short- to super long-range structures.

\end{abstract} 

%\pacs{11.27.+d}
\maketitle

\section{Introduction}
Scalar field models are of current interest in Physics due to their many applications. In particular, they can be used in the investigation of localized structures, such as kinks and lumps \cite{manton,vachaspati}. Kinks are the simplest structures; they are stable due to their topological nature. Considering the canonical Lagrangian density, which consists of the difference between a kinetic term with the derivatives of the field and a potential term, they arise as static solutions of the equation of motion which connect neighbor minima of the potential.

Even though kinks arise under the action of a single scalar field, their associated equation of motion is of second order, with the presence of non-linearity introduced by the potential. To obtain first-order equations, one may use the BPS formalism \cite{bogo,ps}, which relies on the energy minimization, leading to stable configurations. One may also show that solutions obeying the BPS bound engender null stress, satisfying the Derrick's rescaling argument and ensuring stability against contractions and dilations \cite{trilogia1,derrick,hobart}.

In the study of kinks, a well-known model is the sine-Gordon \cite{sine1,sine2}, which supports analytical solutions. It has an interesting feature: its solutions are true solitons, in the sense that they are integrable and persist after collisions \cite{sine3,sine4}. Since kinks can be applied in the braneworld scenario with an extra dimension of infinite extent \cite{rs2,mirjam,dewolfe}, the sine-Gordon model was used in this context to get braneworlds fully described by analytical functions, including the stability of the gravity sector \cite{greemsinegordon,sinebrana2}.

The kink solutions of the sine-Gordon model have exponential tails, that is, they attain their boundary values very fast, so we shall call them short range. The asymptotic behavior of the solutions is very important in the study of collisions, as it is associated with the interactions between the structures. Therefore, studying solutions with distinct tails is of interest. In this direction, one may seek for field profiles with long range. To achieve this, one may consider the limit of the parameter of the double sine-Gordon model \cite{long0} which leads to a potential with null second derivative at the minima; in this case, the solutions engender the long-range character described by power-law tails. Since these solutions extend farther than the usual sine-Gordon ones, their interactions are stronger, so they are also called highly interactive. Over the years, several works dealing with long-range structures have appeared in the literature \cite{long0new,long1new,long2new,long3new,long1,long2,long3,forcelong,forcelong2,long4,long5,long6,long67,long7}.

Recently, another way to modify the kink profile has been studied. Instead of changing only the potential of the field, one may consider the inclusion of other scalar fields that modify the kinetic terms in the Lagrangian density \cite{constr1,constr2,constr3,constr4,constr5,constr6,constr7}. In these models, the extra fields may be used to modify the internal structure and/or the tail of the solutions. The mechanism works within the BPS formalism, which allows for the presence of minimum energy stable solutions that obeys first-order equations. Interestingly, under specific conditions, some equations can be decoupled and one can show that this impacts the geometry of the kink solution. In the case where the kink is modified in its core, one may obtain results similarly to magnetic domain walls in constrained geometries \cite{constrgeom}.

In this manuscript, we investigate the presence of kink structures with logarithmic tails, whose falloff is even slower than the long-range ones with power-law asymptotic behavior. We call them super long-range structures. In Sec.~\ref{sec2}, we investigate how these structures arise under the action of models with a single real scalar field. We associate the super long-range character to the derivatives of the potential calculated in its minima. We introduce a model in which all orders of the derivatives of the potential vanish at the minimum connected by the tail of the kink and obtain an analytical solution with logarithmic tails. The energy density is calculated, the linear stability is also investigated and a discussion about the forces between pairs formed by super long-range kink-kink and kink-antikink is made. In Sec.~\ref{sec3}, we consider the inclusion of an additional scalar field that modifies the kinetic term of the other field. We use the BPS formalism and explore the situation in which one of the fields can be calculated independently from the other. Within this framework, we show how to smoothly go from long- to super long-range or from short- to super long-range configurations. We end the investigation in Sec.~\ref{sec4}, where we present our final remarks and perspectives for future research.

\section{Single scalar field model}\label{sec2}
We consider the canonical action associated to a single real scalar field in $(1,1)$ spacetime dimensions,
\be\label{action}
S = \int dx \,dt\left(\frac12\p_\mu\phi\p^\mu\phi -V(\phi)\right),
\ee
where $V(\phi)$ is the potential. The equation of motion is
\be\label{eom}
\frac{\p^2\phi}{\p t^2}-\frac{\p^2\phi}{\p x^2} +\frac{dV}{d\phi} = 0.
\ee
It is well known that static solutions connecting two neighbor minima of the potential, i.e., $\phi(\pm\infty)=v_\pm$, obey $d^2\phi/dx^2 = dV/d\phi$, with energy density
\be\label{dens}
\rho = \frac12\left(\frac{\p\phi}{\p x}\right)^2 +V(\phi).
\ee
In Ref.~\cite{bogo}, it was shown that non-negative potentials support stable minimum-energy solutions obeying the first-order equation
\be\label{fo}
\frac{d\phi}{dx} = \pm\sqrt{2V(\phi)},
\ee
in which the expressions with upper and lower signs, which describe the increasing (kink) and decreasing (anti-kink) solutions, respectively, are related by $x\to-x$. The solutions of the above equation engender null stress and satisfy Derrick's scaling argument \cite{trilogia1,derrick,hobart}, avoiding instabilities due to contractions and dilations.

One can define the classical mass associated to a minimum $\phi=v_i$ of the potential as
\be\label{massa}
m_{v_i}^2 = \left.\frac{d^2V}{d\phi^2}\right|_{\phi=v_i}.
\ee
The equation of motion allows one to show that, for finite non-null $m_{v_i}$, the asymptotic behavior is given by $v_i-\phi(x)\propto e^{-m_{v_i}|x|}$. As we have previously commented, this occurs in the sine-Gordon model, whose potential is
\be\label{vsg1}
V(\phi) = \frac{1}{2}\cos^2(\phi).
\ee
It supports a family of degenerated minima located at $v_i=(i-1/2)\pi$, with $i\in \mathbb{Z}$. By using Eq.~\eqref{massa}, one can show that the classical mass is the same in all the aforementioned minima, with the value $m^2=1$. Of course, the set of minima allows for the presence of a family of solutions. In the central sector, $\phi\in[-\pi/2,\pi/2]$, we have the kink given by
\be\label{solsg}
\phi(x) = \arcsin(\tanh(x)),
\ee
where we have used $\phi(0)=0$ to fix the constant of integration. This solution is odd. Far away from the origin, for $x\to\infty$ it behaves as $\phi(x)\approx \pi/2-2e^{-x}$, with the expected exponential tail, so we call it short-range solution. The energy density is obtained from Eq.~\eqref{dens}, which reads
\be
\rho(x) = \sech^2(x).
\ee
By integrating it, we get the energy $E=2$.

The exponential tail in the solutions of the sine-Gordon model appears due to the finite non-null character of the classical mass. Since we are interested in obtaining the super long-range structures, let us first review the known long-range solutions. They appear in potentials that support null classical mass. In this direction, one may consider the double sine-Gordon model \cite{long0}, with $V(\phi)=(4/(1+4|\eta|))(\cos(\phi/2)-\eta\cos(\phi))$. One may take the special case $\eta=-1/4$ and perform the change $\phi\to4\phi$ to get the potential $V(\phi)=4\cos^4(\phi)-3/2$. Inspired by this, we write
\be\label{vcos4}
V(\phi) = \frac{1}{2}\cos^4(\phi).
\ee
Albeit this potential is similar to the one in Eq.~\eqref{vsg1}, it has different properties; it was investigated in Ref.~\cite{long1}. Its minima is located at the same points of the usual sine-Gordon potential, but now engender null classical mass. In the central sector, one gets
\be\label{solcos4}
\phi(x) = \arctan(x).
\ee
By expanding it asymptotically, one has $\phi(x)\approx \pi/2-1/x$. Therefore, the tail is of a power-law type, with falloff slower than the exponential one. This is the reason to call these structures long range or highly interactive. Other solutions with power-law tails can be found in the literature \cite{long2,long3,long67}. The energy density \eqref{dens} takes the form
\be\label{rhocos4}
\rho(x) = \frac{1}{\left(1+x^2\right)^2},
\ee
such that $E=\pi/2$.

\begin{figure}[t!]
\centering
\includegraphics[width=0.8\linewidth]{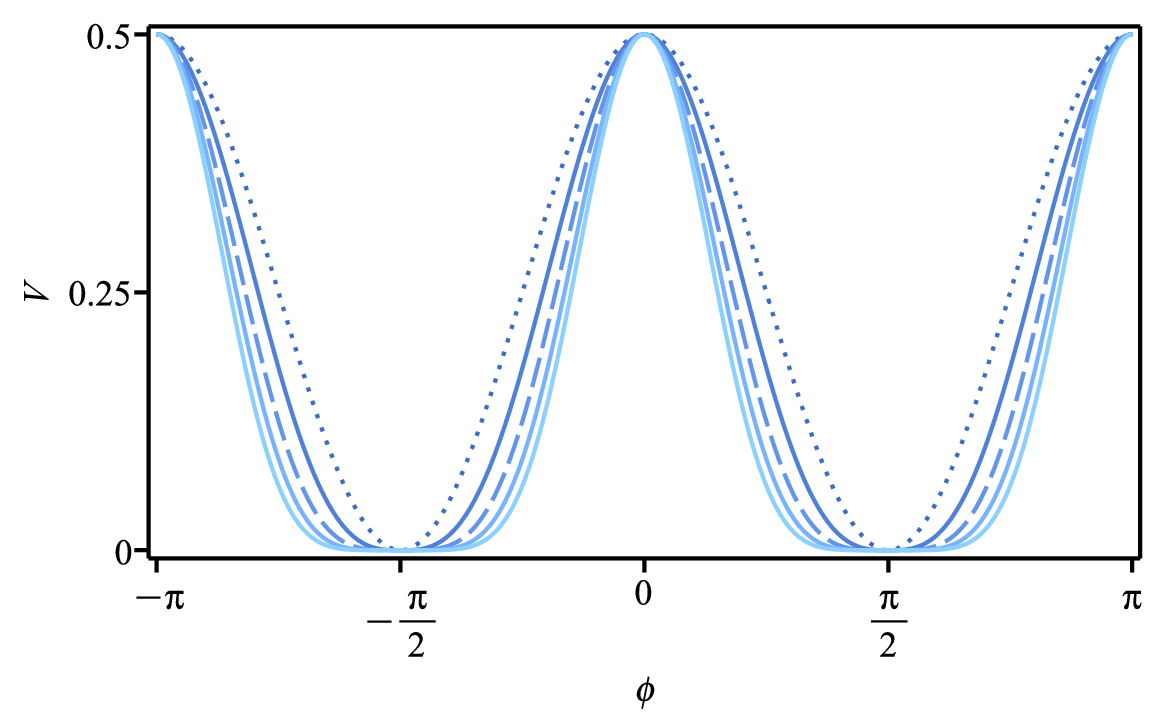}
\includegraphics[width=0.8\linewidth]{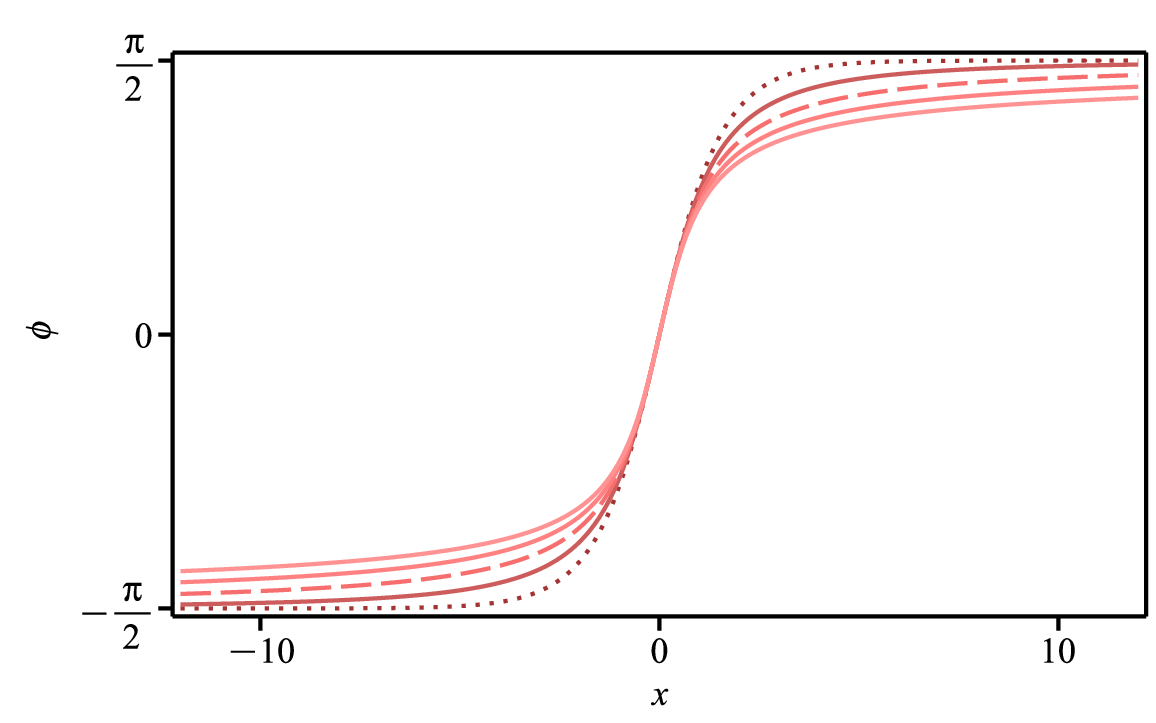}
\caption{The potential in Eq.~\eqref{vcosn} (top, blue colors) and the solution of the first-order equation \eqref{eqn} (bottom, red colors) for the central sector of the potential, for $n=2,3,4,5$ and $6$. The dotted lines represent the case $n=2$, whose potential is in Eq.~\eqref{vsg1} and solution, which engenders exponential tails, is in Eq.~\eqref{solsg}. The dashed lines stand for $n=4$, with potential \eqref{vcos4} and solution \eqref{solcos4} engendering power-law tails. In each panel, the colors get lighter as $n$ increases.}
\label{figcosn}
\end{figure}

We then further investigate the impact of null classical masses and higher derivatives of the potential in the asymptotic behavior of the solutions. In the specific case of Eq.~\eqref{vcos4}, one can show that all the derivatives until third order vanish at the minima $\phi=v_i$. We then consider the potential introduced in Ref.~\cite{long1new} and exchange the exponent of the sine-Gordon model, including a parameter $n$,
\be\label{vcosn}
V(\phi) = \frac{1}{2}\left|\cos^n(\phi)\right|,
\ee
where $n$ is a real parameter which obeys $n>2$. Regardless the value of $n$, the above expression supports minima at the very same location of the usual sine-Gordon potential. However, the derivatives at these points now obey $d^{k}V/d\phi^{k}\big|_{\phi=v_i}=0$ for $k=0,\ldots,\lceil{n-1}\rceil$. Thus, $n$ controls the orders of derivatives which vanish at the minima. Notice that $k$ is integer, with maximum value determined by the ceiling function of argument $n-1$, which we have denoted by $\lceil{n-1}\rceil$. In this situation, the first-order equation \eqref{fo} with upper sign reads
\be\label{eqn}
\frac{d\phi}{dx} =  \big|\cos(\phi)\big|^{n/2}
\ee
This equation cannot be solved analytically for a general $n$. However, before obtaining the solutions, we can analyze the asymptotic behavior. For simplicity, we focus on the central sector, $\phi\in[-\pi/2,\pi/2]$. 
By using it, one can show that the right tail ($x\to\infty$) of the solution obeys $\phi(x)\approx \pi/2-\big((n-2)x/2\big)^{-2/(n-2)}$. Therefore, the parameter $n$ controls how farther the tail extends. As $n$ gets larger and larger, the falloff becomes slower and slower. We then see that the more the order of vanishing derivatives increases, the farther the solution goes. In Fig.~\ref{figcosn}, we display the potential \eqref{vcosn} and we use numerical procedures with the condition $\phi(0)=0$ to show the solution of Eq.~\eqref{eqn} in the central sector of the potential \eqref{vcosn} for several values of $n$.
%%%%%%%%%%%%%%%%%%%%%%%%%%%%%%
%%%%%%%%%%%%%%%%%%%%%%%%%%%%%%

\begin{figure}[t!]
\centering
\includegraphics[width=0.8\linewidth]{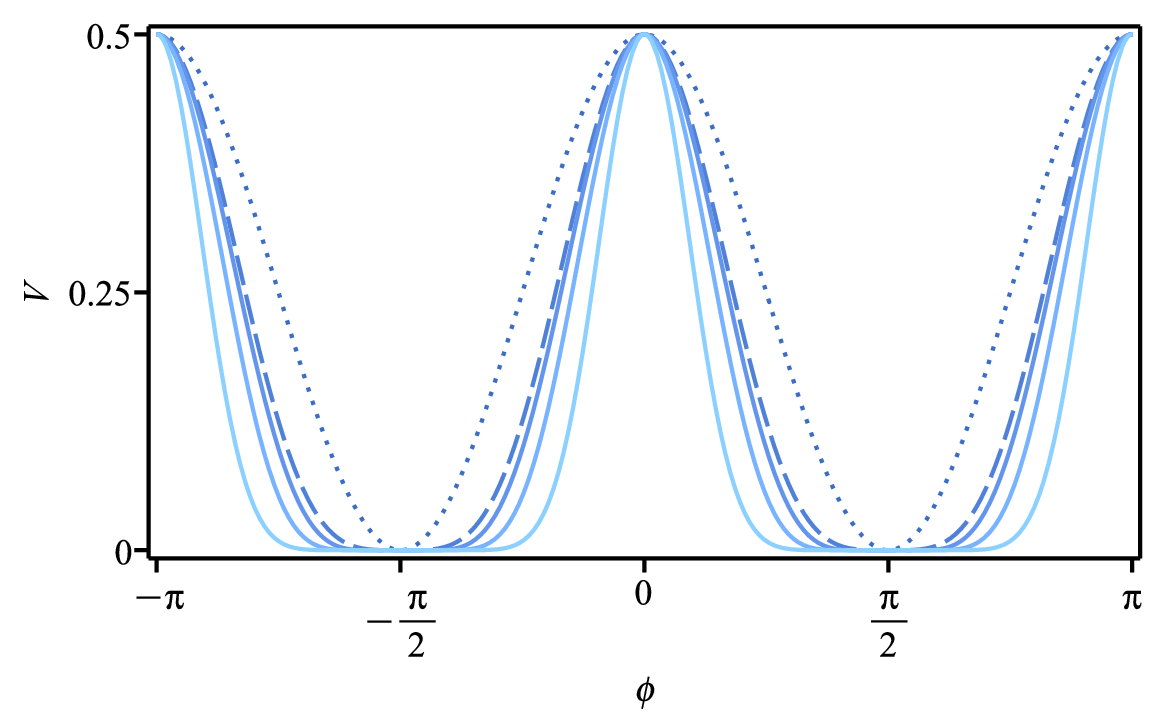}
\includegraphics[width=0.8\linewidth]{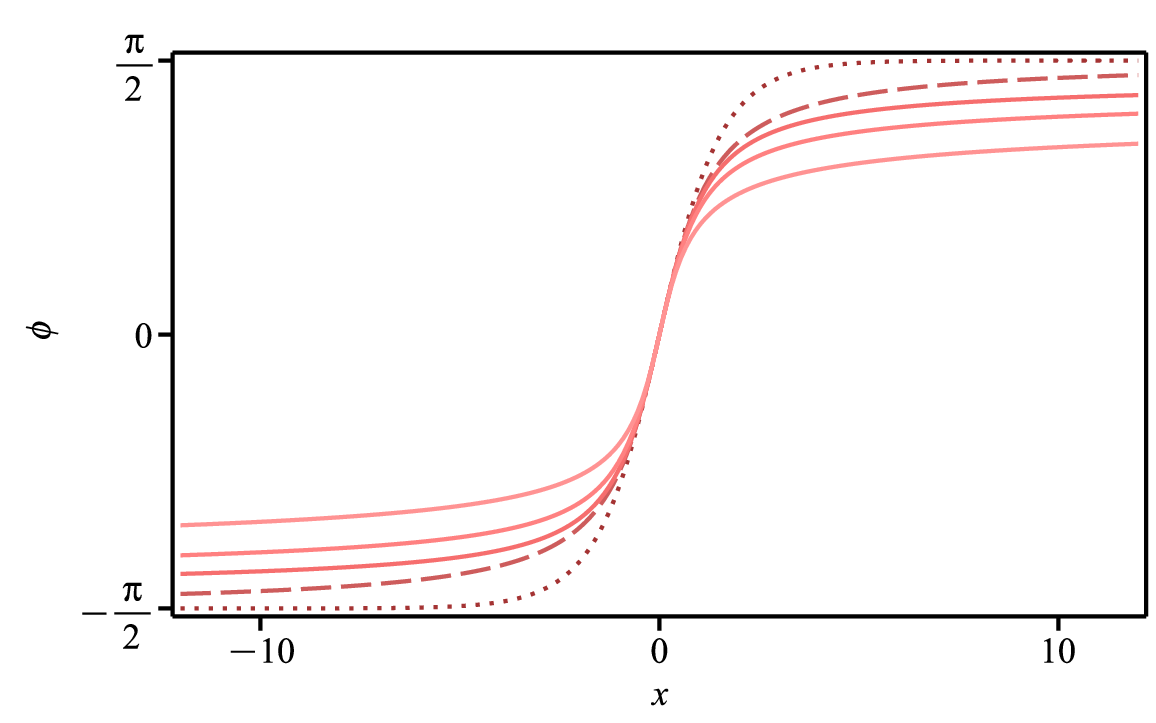}
\caption{The potential \eqref{vsuper} (top, blue colors), the solution \eqref{solsuper} (bottom, red colors) for $a=0,1/2,1$ and $2$. The dotted lines represent the sine-Gordon model in Eqs.~\eqref{vsg1} and \eqref{solsg}, and the dashed lines stand for the case $a=0$, described by \eqref{vcos4} with solution \eqref{solcos4}. In each panel, the colors get lighter as $a$ increases.}
\label{fig2}
\end{figure}

The inclusion of the parameter $n$ allows us to modify the derivatives of the potential and, as a consequence, the tail of the solutions. Notice, however, that what happens in the limit whose derivatives of all orders vanish at the minima, $n\to\infty$, is not clear. To investigate this issue, we consider the model described by the potential
\be\label{vsuper}
V(\phi) = \frac{1}{2}\cos^4(\phi)\,\sech^2(a\tan(\phi))
\ee
and $V(v_i)=0$, where $v_i=(i-1/2)\pi$, with $i\in \mathbb{Z}$. Here, $a$ is a non-negative parameter. The case $a=0$ recovers the potential in Eq.~\eqref{vcos4}. For general $a$, the classical masses \eqref{massa} vanish at all the minima. Furthermore, for $a>0$, \emph{all orders of the derivatives vanish at the minima}, i.e., $d^{k}V/d\phi^{k}\big|_{\phi=v_i}=0$ for $k\in\mathbb{N}$. Even so, the model supports analytical solutions. One can show that the solution of the first-order equation \eqref{fo} with upper sign for the central sector, $\phi\in[-\pi/2,\pi/2]$, of the above potential is
\be\label{solsuper}
\phi(x) = \arctan\left(\frac{\arcsinh(ax)}{a}\right),
\ee
where the constant of integration was fixed as in the previous models ($\phi(0)=0$). As expected, the limit $a\to0$ leads to the solution \eqref{solcos4}, which engenders power-law tails. The above solution obeys $\phi(-x)=-\phi(x)$, being odd as the previous ones. In the case $a>0$, the asymptotic analysis ($x\to\infty$) of this kink structure leads to
\be\label{solsuperasy}
\begin{aligned}
\phi(x) &= \frac{\pi}{2} -\frac{a}{\ln(2ax)} +\frac{a^3}{3\ln^3(2ax)} +\frac{1}{4x^2\ln^2(2ax)}\\
    &+\mathcal{O}\left(\frac{1}{\ln^5(2ax)}\right) + \mathcal{O}\left(\frac{1}{x^2\ln^4(2ax)}\right).
\end{aligned}
\ee
The presence of the parameter $a$ is not trivial in the above expression. Indeed, by comparing the third and fourth terms, we see that they may have similar weights for a given interval of $x$, depending on $a$. We have shown that the aforementioned terms are equal for $x=e^{-(1/2)W_{-1}(-2a^2/3)}/(2a)$, where $W_{-1}(y)$ is a branch of the Lambert function, with $-e^{-1}\leq y<0$. Therefore, this can only occur for $a\leq\big(\sqrt{6}/2\big)e^{-1/2}\approx 0.743$. The above expansion possesses a mix between purely-logarithmic and logarithmic-power-law terms. Nevertheless, for very large distances from the origin, scaled by $x\gg e^{a}/(2a)$, we can use the approximation
\be\label{phiasyln}
\phi_{asy}(x) \approx \frac{\pi}{2} -\frac{a}{\ln\big(2ax\big)}.
\ee
We then see that the solution \eqref{solsuper} engenders logarithmic tails, falling off even slower than the so-called long-range structures. Due to this feature, we call the expression in \eqref{solsuper} \emph{super long-range} solution. In Fig.~\ref{fig2}, we display the potential \eqref{vsuper} and the super long-range solution \eqref{solsuper} for some values of the parameters. Notice that, as $a$ gets larger, the concavity of the potential at the minima becomes wider and the tails of the solutions go farther as a consequence of the above logarithmic decay.

\begin{figure}[t!]
\centering
\includegraphics[width=0.8\linewidth]{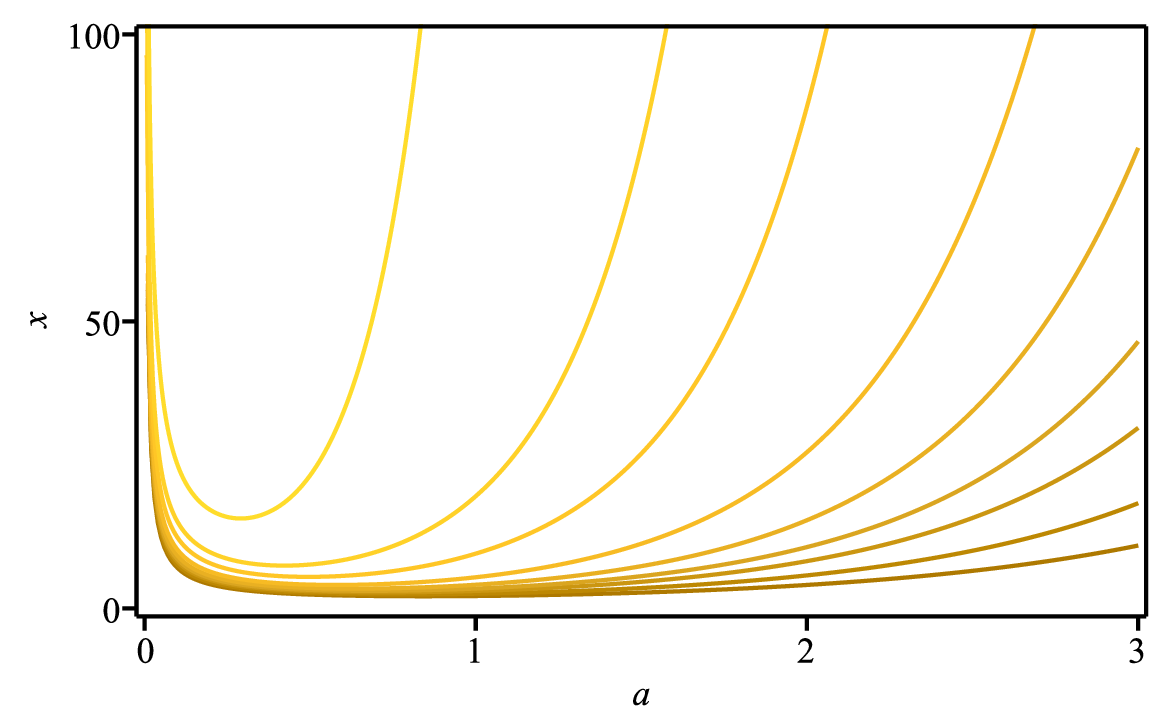}
\caption{The values of $x$ in terms of $a$ which satisfy the equation $\phi_{asy}(x)/\phi(x)=\epsilon$ for $\epsilon=0.9,0.93,0.95,0.96,0.97,$ $0.98,0.99,0.995$ and $0.999$. The colors get lighter as $\epsilon$ approaches $1$.}
\label{fig33}
\end{figure}

To show how far one must be from the origin to use the approximation in Eq.~\eqref{phiasyln}, we plot, in Fig.~\ref{fig33}, the values of $x$ with respect to $a$ which satisfy the ratio $\phi_{asy}(x)/\phi(x)=\epsilon$, where $\epsilon$ must be near $1$. This is done to show how good the approximation \eqref{phiasyln} with respect to $a$ is. Notice that, for some specific value of $a$, one must go farther and farther away from the origin as $\epsilon$ approaches $1$, exhibiting the super long-range nature of the solution \eqref{solsuper}. Also, for very small values of $a$, the asymptotic behavior \eqref{phiasyln} is only valid for $x\approx1/(2a)$.
%%%%%%%%%%%%%%%%%%%%%%%%%%%%%%
%%%%%%%%%%%%%%%%%%%%%%%%%%%%%%

%%%%%%%%%%%%%%%%%%%%%%%%%%%%%%
%%%%%%%%%%%%%%%%%%%%%%%%%%%%%%

The energy density associated to the solution \eqref{solsuper} can be calculated from Eq.~\eqref{dens}, which reads
\be\label{rhosuper}
\rho(x) = \frac{a^4}{\left(1+a^2x^2\right)\big(a^2+\arcsinh^2(ax)\big)^2}.
\ee
Near the origin, it can be approached by $\rho(x\approx0)\approx1-(a^2+2)x^2$. By investigating the asymptotic behavior of the above energy density, we obtain
\be
\begin{aligned}
\rho(x)&= \frac{a^2}{x^2\big(a^2+\ln^2(2ax)\big)^2} -\frac{1}{x^4}\Bigg(\frac{1}{\big(a^2+\ln^2(2ax)\big)^2}\\
    &+\frac{\ln(2ax)}{\big(a^2+\ln^2(2ax)\big)^3}\Bigg) +\mathcal{O}\left(\frac{1}{x^6}\right).
\end{aligned}
\ee
Therefore, contrary to the solutions, the energy density always falls off with terms that mix power-law and logarithmic functions. Similarly to the solutions, in the region where $x\gg e^{a}/(2a)$, we can neglect terms of higher order and write
\be
\rho(x)\approx \frac{a^2}{x^2\ln^4(2ax)}.
\ee
The presence of the power-law term in the above expression makes this approximation in the energy density being valid for distances smaller than the one required by the solution \eqref{solsuperasy}, which has a pure-logarithmic asymptotic behavior. We remark that, even though this falloff of the energy density is slower than the usual power-law one, it is still faster than the one related to the vacuumless model, whose solutions are unlimited and arise from potentials with runaway minima, investigated in Refs.~\cite{vacuum1,vacuum2}. However, we note that, contrary to the vacuumless configurations, the solution \eqref{solsuper} emerges with the potential \eqref{vsuper}, which is somewhat similar to the well-known sine-Gordon model, so the asymptotic values of the solution are finite. 
%%%%%%%%%%%%%%%%%%%%%%%%%%%%%%
\begin{figure}[t!]
\centering
\includegraphics[width=0.8\linewidth]{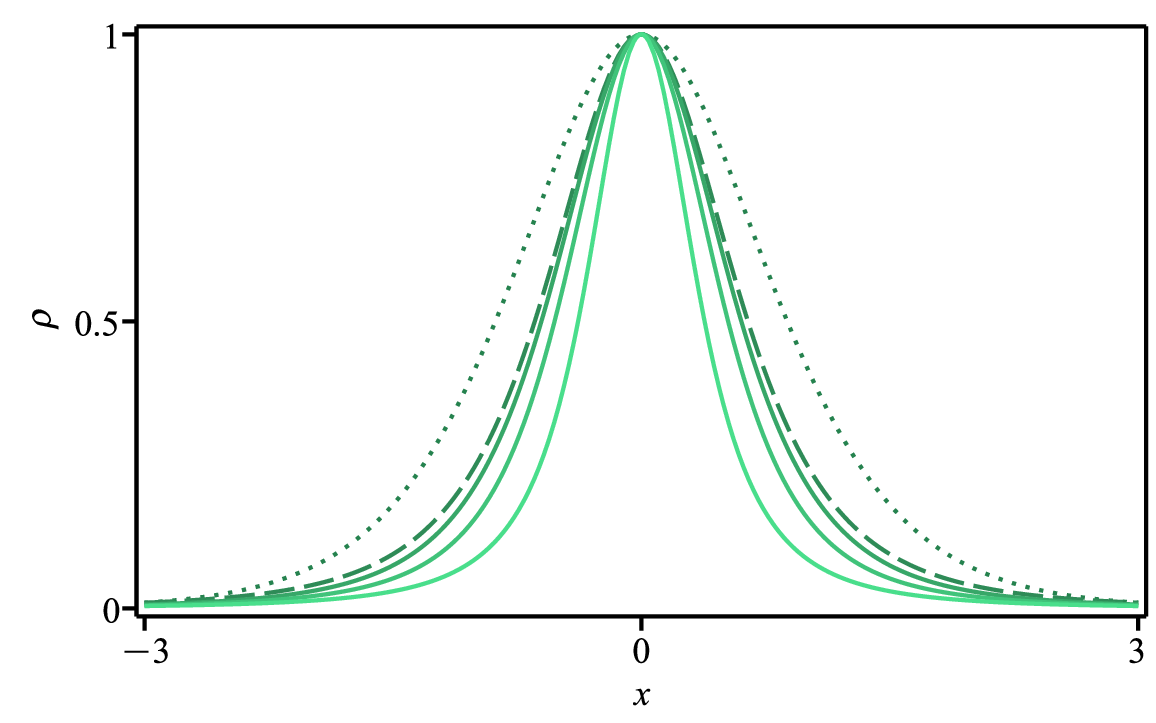}
\includegraphics[width=0.8\linewidth]{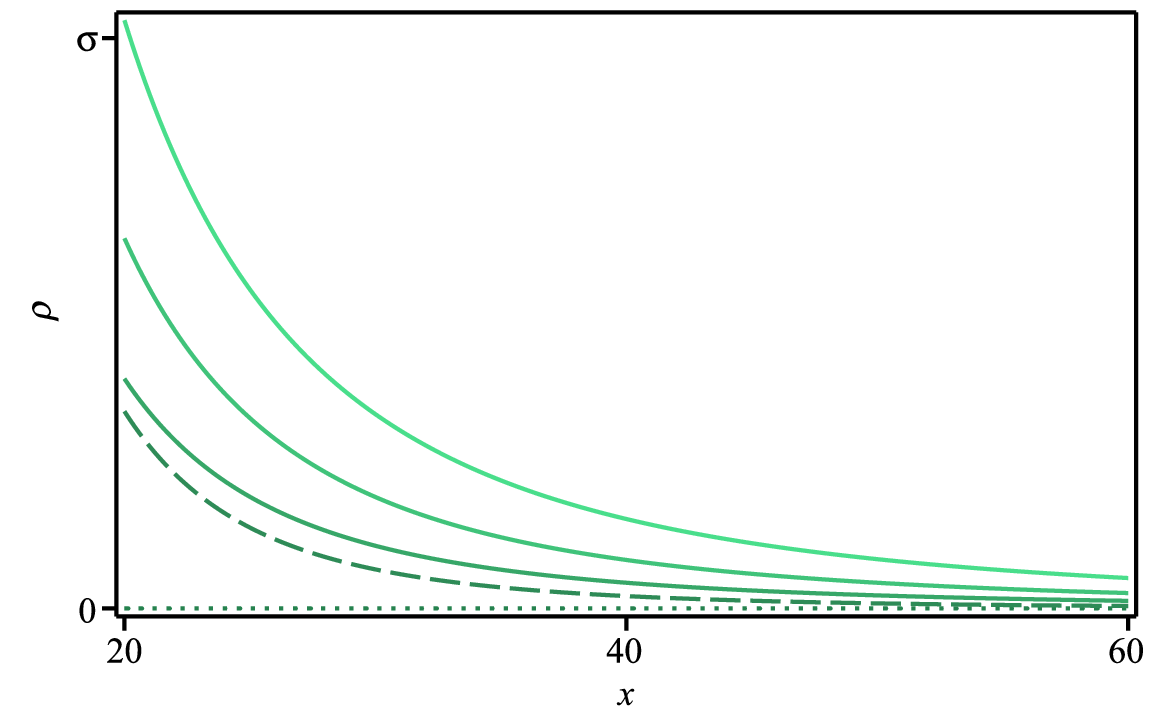}
\caption{The energy density \eqref{rhosuper} for the same line styles and values of the parameters used in Fig.~\ref{fig2}. The top panel shows its general behavior and the bottom panel depicts its asymptotic behavior with $x$; the scale of the vertical axis is defined by $\sigma=1.8\times 10^{-5}$. The green colors get lighter as $a$ increases.}
\label{fig3}
\end{figure}
%%%%%%%%%%%%%%%%%%%%%%%%%%%%%%
%%%%%%%%%%%%%%%%%%%%%%%%%%%%%%
\begin{figure}[t!]
\centering
\includegraphics[width=0.8\linewidth]{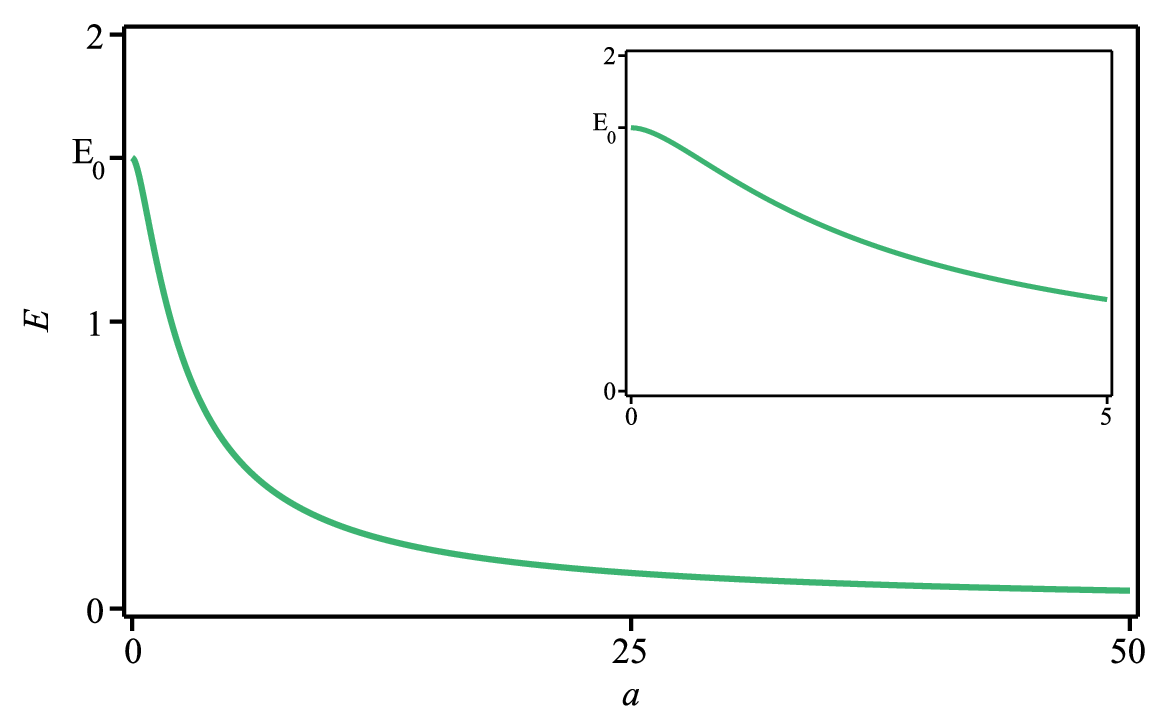}
\caption{The behavior of the energy with respect to $a$. The inset displays the behavior for small values of $a$, showing that $E_0\to \pi/2$ as $a\to0$.}
\label{fig4}
\end{figure}
%%%%%%%%%%%%%%%%%%%%%%%%%%%%%%
In Fig.~\ref{fig3}, we show the plot of the energy density \eqref{rhosuper} for the same parameters used in Fig.~\ref{fig2}. It is worth commenting that, even though the falloff of the energy density is faster with increasing $a$ in regions near the origin, it actually tends to become slower as we go away from the origin, as expected from the behavior of the solutions.

The slow falloff of the super long-range configuration gives rise to a discussion about the thickness of the solution. As it is known in the literature, there are some ways to define thickness. One of them is using the energy density. Since the energy density vanishes asymptotically, we can define the thickness as $\Delta=2x_t$, where $x_t$ is the point in which the energy density obeys $\rho(x_t)=\zeta$, with $\zeta$ being some value of interest near zero. Another way is by using some point $\tilde{x}$, such that $E_{\text{inside}}/E = \vartheta$, where $\vartheta$ represents the percentage of the total energy which is inside the interval $[-\tilde{x},\tilde{x}]$. In this situation, the thickness is $\Delta=2\tilde{x}$.

The energy density \eqref{rhosuper} cannot be integrated analytically, so we have used numerical procedures to get the plot of the energy in Fig.~\ref{fig4}; the energy is decreasing with $a$, showing that solutions with longer tails engender lesser energies. In the limit $a\to0$, we get $E_0=\pi/2$, matching with the expected value obtained below Eq.~\eqref{rhocos4}, since $a=0$ recovers the potential \eqref{vcos4}.

\subsection{Linear stability}
Since the solution \eqref{solsuper} engenders very long tails, let us examine its stability in the presence of small fluctuations. To do so, we consider $\phi(x,t)=\phi_s(x)+\sum_n\psi_n(x)\cos(\omega_nt)$, where $\phi_s(x)$ is the static solution of Eq.~\eqref{fo}. By substituting this time-dependent field in the equation of motion \eqref{eom}, we get the stability equation
\be\label{stabeq}
   -\frac{d^2\psi_n}{dx^2} +U(x)\psi_n = \omega_n^2\psi_n,\quad U(x)=d^2V/d\phi^2\big|_{\phi=\phi_s(x)},
\ee
in which $U(x)$ is the stability potential. This is a one-dimensional Schr\"odinger-like eigenvalue equation with zero mode ($\omega=0$) \cite{modozero1,modozero2}
\be\label{psi0}
\psi_0 = \frac{1}{\sqrt{E}}\frac{d\phi_s}{dx},
\ee
where $E$ is the energy of the static solution. The linear stability is ensured if negative eigenvalues are absent in the stability equation. This occurs if the above zero mode does not present nodes.

The stability potential in \eqref{stabeq} associated to the sine-Gordon model \eqref{vsg1} takes the form of a modified P\"oschl-Teller,
\be\label{usg}
U(x)=1-2\,\sech^2(x),
\ee
with the zero mode being the only bound state; it also supports a semi-bound state with $\omega^2=1$. For the potential \eqref{vcos4}, which leads to the usual long-range structures (power-law tails), we have 
\be\label{ucos4}
U(x)=\frac{6x^2-2}{\big(1+x^2\big)^2},
\ee
which has a volcano shape, vanishing asymptotically; there is a single bound state in this case: the zero mode.

For the super long-range solution \eqref{solsuper}, the stability potential in \eqref{stabeq} is
\be\label{usuper}
\begin{aligned}
U(x) &= \frac{2a^2\big(3\,\arcsinh^2(ax)-a^2\big)}{\big(1+a^2x^2\big)\big(a^2+\arcsinh^2(ax)\big)^2}-\frac{a^2\big(1-2a^2x\big)}{1+a^2x^2}\\    &+\frac{6a^3x\,\arcsinh(ax)}{\left(1+a^2x^2\right)^{3/2}\big(a^2+\arcsinh^2(ax)\big)}.
\end{aligned}
\ee
The asymptotic behavior of the above potential is given by
\be
\begin{aligned}
U(x) &= \frac{2}{x^2}\left(1+\frac{3\ln^2(2ax)}{a^2+\ln^2(2ax)}+\frac{3\ln^2(2ax)-a^2}{\big(a^2+\ln^2(2ax)\big)^2}\right)\\
    &-\frac{1}{a^2x^4}\Bigg(5 +\frac{18\ln^2(2ax)-3}{2\big(a^2+\ln^2(2ax)\big)}\\
    &+\frac{3\ln(2ax)\big(3\ln(2ax)-1\big)-2a^2
    }{\big(a^2+\ln^2(2ax)\big)^2} \\
    &+\frac{2\ln(2ax)\big(3\ln^2(2ax)-a^2\big)
    }{\big(a^2+\ln^2(2ax)\big)^3}\Bigg) +\mathcal{O}\left(\frac{1}{x^6}\right),
\end{aligned}
\ee
exhibiting combinations of power-law and logarithmic functions. For very large distances from the origin, $x\gg e^{a}/(2a)$, we can disregard terms of higher order and take $U(x) \approx 2/x^2$.

\begin{figure}[t!]
\centering
\includegraphics[width=0.75\linewidth]{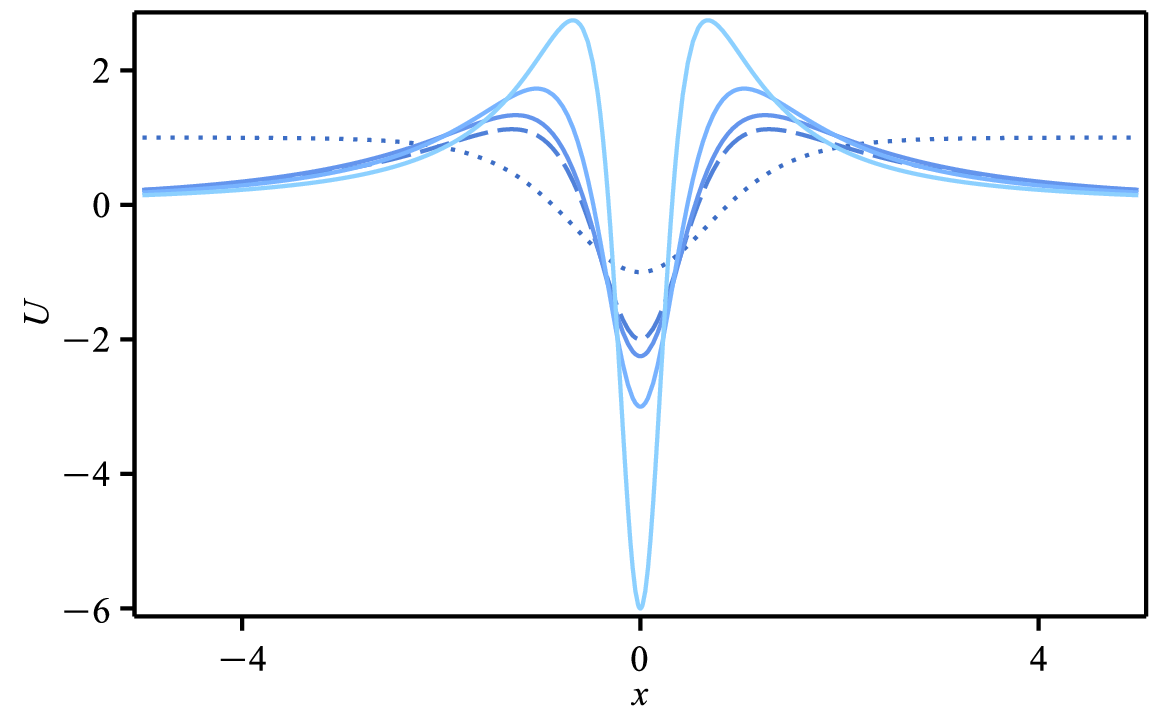}
\caption{The stability potential \eqref{usuper} for the same values of the parameter $a$ in Fig.~\ref{fig2}. The dotted lines represent the sine-Gordon case \eqref{usg} and dashed lines stand for the case $a=0$, described by \eqref{ucos4}. The blue colors get lighter as $a$ increases.}
\label{fig5}
\end{figure}

The zero mode \eqref{psi0} associated to \eqref{solsuper} is given by
\be
\psi_0(x) = \frac{1}{\sqrt{E}}\frac{a^2}{\sqrt{1+a^2x^2}\big(a^2+\arcsinh^2(ax)\big)}.
\ee
It does not engender nodes for all $a$. Therefore, the super long-range solution \eqref{solsuper} is linearly stable. In Fig.~\ref{fig5}, we display the potential \eqref{usuper} for some values of the parameters. 
%%%%%%%%%%%%%%%%%%%%%%%%%%%%%%
%%%%%%%%%%%%%%%%%%%%%%%%%%%%%%

\subsection{Interkink forces}
As it is well known, the asymptotic behavior of the solutions is of great importance in the study of their interactions. In the usual long-range kinks, the kink-kink force is repulsive and the kink-antikink force is attractive; they decay with the fourth power of the separation between the structures \cite{forcelong}. To calculate the force between super long-range solutions, we follow the lines of Ref.~\cite{force}. The momentum $P(t)$ in the region $[x_1,x_2]$ associated to the action \eqref{action} is calculated standardly; it is given by $P(t) = -\int_{x_1}^{x_2} dx\,(\p\phi/\p t) (\p\phi/\p x)$. This allows us to calculate the force, $F=d P/d t$. By combining these expressions with the equation of motion \eqref{eom}, we get
\be
F = \left[-\frac12\left(\frac{\p\phi}{\p t}\right)^2 -\frac12\left(\frac{\p\phi}{\p x}\right)^2 +V(\phi)\right]_{x_1}^{x_2}.
\ee
Since we are interested in calculating the forces between super long-range solutions, we have two configurations possible: kink-antikink ($KA$) and kink-kink ($KK$). Let us consider that the structures are symmetrically located around the origin, each one at the distance $L$ from the point $x=0$, where $L$ is much larger than the thickness of the solutions. In this situation, the separation is $2L$. Supposing that the structures are initially ($t=0$) at rest, the force in the kink at the left side is
\be\label{forcegen}
F = \left[-\frac12\left(\frac{\p\phi}{\p x}\right)^2 +V(\phi)\right]^0_{-\infty}.
\ee

We describe the pairs $KA$ and $KK$ with an approximation
\bal\label{phika}
\phi^{KA} &= \phi_l^K(x+L) +\phi_l^A(x-L) -\phi_l^K(\infty),\\ \label{phikk}
\phi^{KK} &= \phi_l^K(x+L) +\phi_{l+1}^K(x-L) -\phi_l^K(\infty),
\eal
where $\phi_l(x)$ is a solution of the equation \eqref{fo} and the index $l$ represent some sector of the potential. Therefore, the $KA$ configuration involves solutions of the same sector ($\phi_l$) while the $KK$ pair must take into account solutions of neighbor sectors ($\phi_l$ and $\phi_{l+1}$).

For the sine-Gordon solutions \eqref{solsg}, we get $F=\pm 8e^{-2L}$, being positive (attractive) for the pair $KA$ and negative (repulsive) for configuration $KK$ \cite{forcesine}. Notice that the force decays with an exponential function of the separation. For the specific long-range solution in \eqref{solcos4}, we get the attractive force $F=8L^{-4}$ for the $KA$ pair and the repulsive force $F=-2L^{-4}$ for the $KK$ configuration. Therefore, the power-law behavior of the solution also appears in the force. In Ref.~\cite{forcelong}, a comparison of these values with the ones obtained from accelerating kinks was made and, even though the dependence on the distance matches, the coefficient does not. To obtain a better coefficient, a method different from \eqref{phika} and \eqref{phikk} was proposed. We shall refer to it as the gluing technique. It consists of cutting the left kink at $x=-L/2$ and the right kink at $x=L/2$ and gluing them with a function obtained from an approximation with adjusted coefficients of a Taylor expansion of \eqref{phika} and \eqref{phikk} at the origin.

\begin{figure}[t!]
\centering
\includegraphics[width=0.8\linewidth]{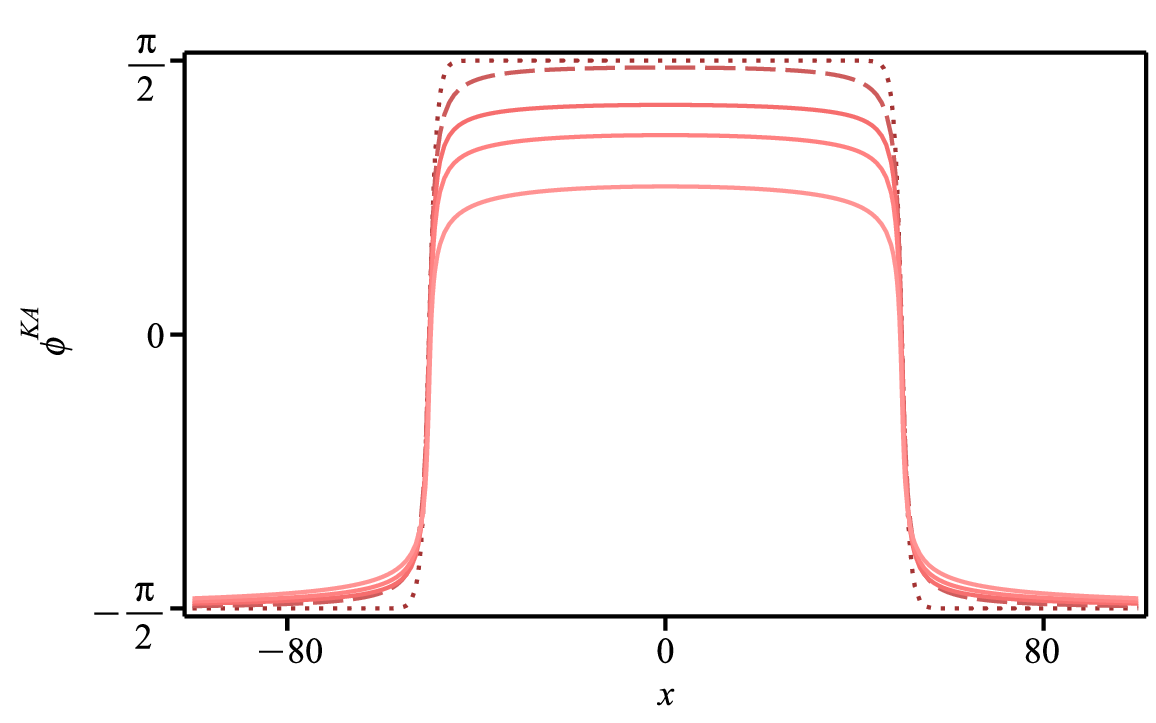}
\includegraphics[width=0.8\linewidth]{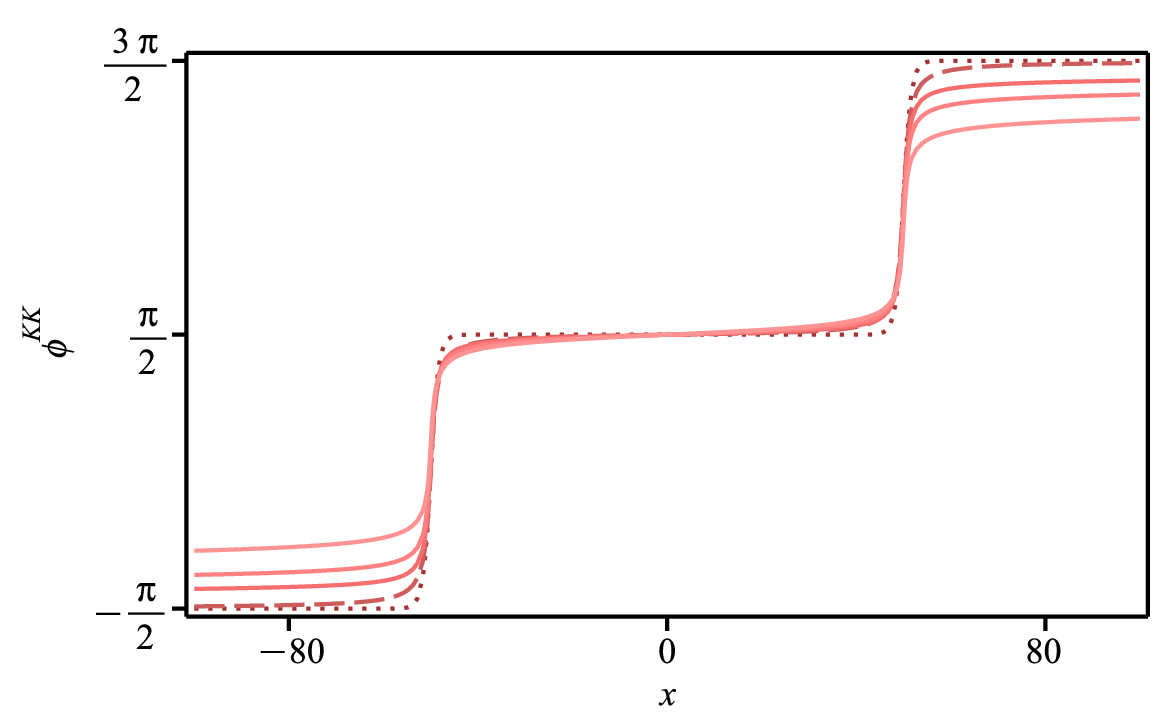}
\caption{The kink-antikink (KA) pair \eqref{phika} (top) and the kink-kink (KK) pair \eqref{phikk} (bottom) for $L=50$ and the same values of $a$ used in Fig.~\ref{fig2}. The line styles also follow the aforementioned figure.}
\label{fig6}
\end{figure}

In the case of the super long-range solution \eqref{solsuper}, we have shown that tail is logarithmic, extending farther than the power-law and exponential ones. In Fig.~\ref{fig6}, we display the $KA$ pair formed by solutions in the interval $[-\pi/2,\pi/2]$ and the $KK$ configuration formed by a solution in $[-\pi/2,\pi/2]$ joint with a solution in $[\pi/2,3\pi/2]$. Notice that, even though $L=50$ seems to be reasonable for our approximation in the case of exponential and power-law tails, it is not sufficient for the super long-range solutions, because the tails are too far from the minima connected by the solutions. This shows that the falloff of the solution \eqref{solsuper} is indeed \emph{very slow}, so one must consider higher separations to use the approximated solutions \eqref{phika} and \eqref{phikk}. For sufficiently large $L$ and $a>0$, the $KA$ force is
\be\label{fkanaive}
F^{KA} = \frac{16a^3}{L\ln^4(2aL)},
\ee
which decays slower that the forces associated to the configurations with exponential and power-law tails. Notwithstanding that, the above force is attractive, as usually occurs for $KA$ pairs. The $KK$ force is given by
\be\label{fkknaive}
F^{KK} = -\frac{2a^2}{L^2\ln^4(2aL)}.
\ee
We then have an attractive force, as usual. Remarkably, it decays faster than the $KA$ force due to the square $L$ in the denominator. This is compatible with Fig.~\ref{fig6}, where we see that the approximation with $L=50$ works better for the $KK$ than for the $KA$ configuration. Indeed, we can use the behavior of the $KA$ pair near the origin to estimate a good separation. Ideally, this configuration should be in the vacuum at the origin, i.e., $\phi^{KA}(0)=\pi/2$. We have found that, for $a=1/2,1$ and $2$, respectively, the equality $\phi^{KA}(0)=0.9\times\pi/2$ requires $L\approx 574, 1.65\times 10^{5}$ and $2.72\times 10^{10}$; all of them are much larger than $L=50$.
%%%%%%%%%%%%%%%%%%%%%%%%%%%%%%
%%%%%%%%%%%%%%%%%%%%%%%%%%%%%%

To verify the expressions \eqref{fkanaive} and \eqref{fkknaive}, we proceed similarly as in Ref.~\cite{forcelong,forcelong2} and compare them with the results obtained from accelerating kinks. To do so, we consider that the solutions used to form the configurations are well separated at time-dependent positions $-L(t)$ and $L(t)$. Following the aforementioned references, we model the accelerating kink at the left side by
\be
\phi(x,t)=\vphi^K(x+L(t))
\ee
for the same sectors of the potential studied with the approximation \eqref{phika} and \eqref{phikk} plotted in Fig.~\ref{fig6}. Notice that we have neglected the kink at the right side in the above approximation as we shall consider the interval $x\in[-L(t),0]$. By substituting the above field in the equation of motion \eqref{eom}, we get
\be\label{eqaccel}
\frac{d^2\vphi^K}{dy^2} \pm A\frac{d\vphi^K}{dy} -\left.\frac{dV}{d\phi}\right|_{\phi=\vphi^K} = 0,
\ee
in which $y=x+L(t)$. In this expression, we have disregarded the term proportional to $(dL/dt)^2$ and taken the acceleration as $\pm A=-d^2L/dt^2$. The positive/negative sign in \eqref{eqaccel} represents attractive/repulsive force associated with the $KA$/$KK$ configuration. Since the separation is large, we can assume that the acceleration is very small, $A\ll1$. In this regime, we can use the BPS equation \eqref{fo} to substitute $d\vphi^K/dy$ by $dW/d\phi|_{\phi=\vphi^K}$ in the factor multiplying $A$ in \eqref{eqaccel}, in which the auxiliary function $W(\phi)$ was introduced as usual in the BPS formalism \cite{bogo}. From this, we can reduce \eqref{eqaccel} to the first order, in the form
\be\label{foaccel}
\frac{d\vphi^K}{dy} =\sqrt{2V(\vphi^K)\mp 2AW(\vphi^K)}.
\ee
To keep the energy $E=\Delta W=M$, we fix the constant of the auxiliary function with $W(-\pi/2)=0$, so $W(\pi/2)=M$. By substituting the above expression into \eqref{forcegen}, one gets the force proportional to $MA$. Therefore, to obtain the force associated with accelerating kinks, we must integrate Eq.~\eqref{foaccel}. For the super long-range solution \eqref{solsuper}, the potential is \eqref{vsuper} and the tail can be obtained from \eqref{solsuperasy}. As it was done in Ref.~\cite{forcelong}, we take $\vphi^K$ and $\phi$ to diverge at the same points of the extrapolated tails, so we write
\be\label{phiktail}
\frac{\pi}{2} - \vphi^K \approx \frac{a}{\ln(2a(x+L))}.
\ee
Since we are working within the region of the super long-range tail, we can take approximated forms for $V(\vphi^K)$ and $W(\vphi^K)$, so \eqref{foaccel} becomes
\be
\frac{d\vphi^K}{dy} =\sqrt{4\left({\vphi^K}-\frac{\pi}{2}\right)^4e^{-\frac{2a}{|{\vphi^K}-\pi/2|}} \mp 2AM},
\ee
where we have used the approximation $W(\vphi^K) = W(\pi/2)=M$ which, in the super long-range structure, is worse than in the cases of power-law and exponential tails. At this point we must be careful to properly choose the configuration before proceeding.

First, we consider the $KA$ pair. We then see that, at $x=-L$ ($y=0$), one has $\vphi^K=-\infty$ (using lateral limit in \eqref{phiktail}) and, at $x=0$ ($y=L$), the derivative of $\vphi^K$ vanishes and the above expression leads us to $\left({\tilde{\vphi}^K}-\pi/2\right)^4e^{-\frac{2a}{|{\tilde{\vphi}^K}-\pi/2|}} = AM/2$. We perform the change of variables $\chi = \vphi^K-\pi/2$ and integrate the above equation to get
\be\label{eqchika}
\int_{-\infty}^{\tilde{\chi}}\frac{d\chi}{\sqrt{4\chi^4e^{-\frac{2a}{|\chi|}} -2AM}} = L,
\ee
where $\tilde{\chi}$ is the real solution of $\tilde{\chi}^4e^{-2a/|\tilde{\chi}|} =AM/2$ compatible with $A\ll1$.
 The above integral is definite, so it leads to an expression in the form $f(A)=L$, from which we get $A=f^{-1}(L)$. Since the force between the $KA$ pair is attractive, we can write $F=Mf^{-1}(L)$. For the $KK$ pair, we can use similar arguments, with $\vphi^K=\pi/2$ at $x=0$, to obtain
\be\label{eqchikk}
\int_{-\infty}^{0}\frac{d\chi}{\sqrt{4\chi^4e^{-\frac{2a}{|\chi|}} +2AM}} = L.
\ee
In this situation, we can write the repulsive force as $F=-Mh^{-1}(L)$, where $A=h^{-1}(L)$ emerges from the above expression.

\begin{figure}[t!]
\centering
\includegraphics[width=0.8\linewidth]{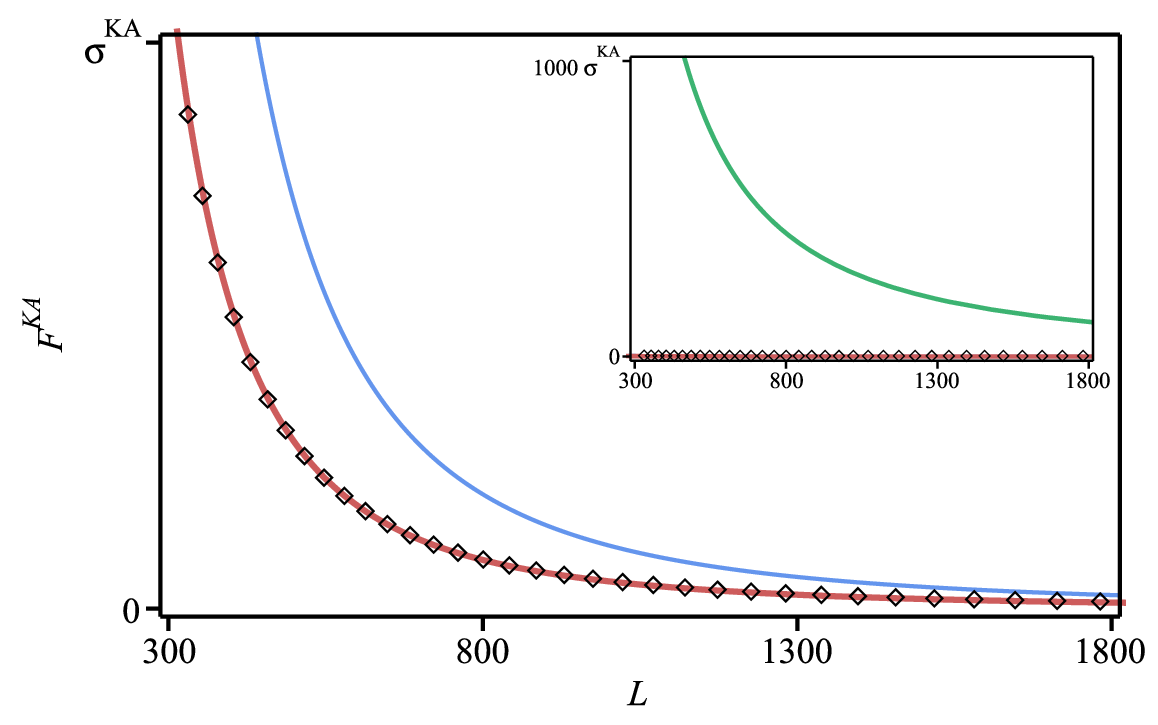}
\includegraphics[width=0.8\linewidth]{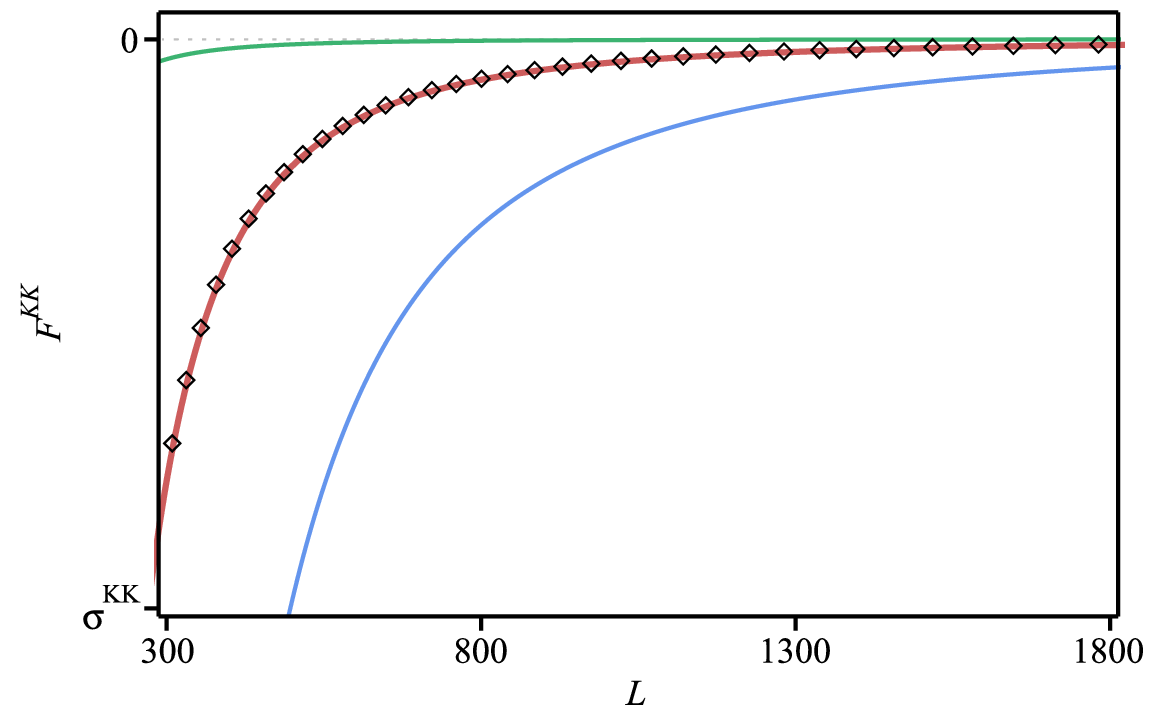}
\caption{The force associated to the kink-antikink ($KA$) pair (top panel) and the kink-kink ($KK$) pair (bottom panel) with respect to the parameter $L$, obtained from three different methods, for $a=1/2$. The solid lines in green stand for the forces \eqref{fkanaive} (top panel) and \eqref{fkknaive} (bottom panel), obtained with the approximation in Eqs.~\eqref{phika} and \eqref{phikk}; there is no green line in the top panel because the force \eqref{fkanaive} is too discrepant from the ones obtained from the other methods. The inset shows the scale needed in the vertical axis to make the green line appear; it is a thousand times larger than the full panel. The diamonds in black stand for the forces obtained numerically from Eqs.~\eqref{eqchika} (top panel) and \eqref{eqchikk} (bottom panel) for $M=1.459$, which are fitted by the curves in red, described by $F^{KA}=0.33/(L^2\ln^2(L))$ and $F^{KK}=-60/(L^2\ln^4(20L))$. The lines in blue denote the forces in Eqs.~\eqref{forceglueka} and \eqref{forcegluekk}, obtained from the gluing method. The scale of the vertical axis is represented by $\sigma^{KA}=3.0\times10^{-9}$ in the top panel and  $\sigma^{KK}=-1.5\times10^{-7}$) in the bottom panel.}
\label{figforcenew}
\end{figure}

Both Eqs.~\eqref{eqchika} and \eqref{eqchikk} cannot be solved analytically. We then use numerical procedures to obtain $A$ and plot the force associated to the $KA$ and $KK$ pairs in the diamond-shaped points in Fig.~\ref{figforcenew}. The curves of the accelerating-kink forces can be compared with the ones obtained with the approximation \eqref{phika} and \eqref{phikk}. Looking at the top panel of Fig.~\ref{figforcenew}, we cannot see the curve representing the force \eqref{fkanaive} because it is too discrepant from values represented by the diamonds. This discrepancy can be seen in the inset.
%%%%%%%%%%%%%%%%%%%%%%%%%%%%%%
%%%%%%%%%%%%%%%%%%%%%%%%%%%%%%

To find some other expression for the force that allows us to compare it with the values from the accelerating kinks or Eqs.~\eqref{phika} and \eqref{phikk}, we use the gluing technique \cite{forcelong} mentioned earlier which, in the region of interaction between the $KA$ pair, leads us to construct the field
\be\label{phikagluing}
\phi^{KA} =
\begin{cases}
\frac{\pi}{2} -\frac{a}{\ln(2a|x+L|)}, &x< -\frac{L}{2}\\
\frac{\pi}{2} -\frac{2a\ln(aL)-a}{2\ln^2(aL)} -\frac{2ax^2}{L^2\ln^2(aL)}, &|x|\leq\frac{L}{2}\\
\frac{\pi}{2} -\frac{a}{\ln(2a|x-L|)}, &x> \frac{L}{2}.
\end{cases}
\ee
In this case, the associated force \eqref{forcegen} is
\be\label{forceglueka}
F^{KA} = \frac{2a^2}{L^2\ln^4(aL)}.
\ee
By comparing it with the expression in Eq.~\eqref{fkanaive}, we see that the coefficients and the exponent of the distance do not match. However, in the top panel of Fig.~\ref{figforcenew}, we can see that the above expression approaches the curve obtained from the accelerating kinks better than \eqref{fkanaive}, which does not appear in the plot because it is too discrepant. Moreover, by adjusting the parameters of the above expression, with $F^{KA}=0.33/(L^2\ln^2(L))$, we were able to get a good fit for the diamonds that describe the accelerating kinks.

For the $KK$ pair, the gluing method leads us to
\be\label{phikkgluing}
\phi^{KK} =
\begin{cases}
\frac{\pi}{2} -\frac{a}{\ln(2a|x+L|)}, &x< -\frac{L}{2}\\
\frac{\pi}{2} +\frac{a(3\ln(aL)-1)x}{L\ln^2(aL)} -\frac{4a(\ln(aL)-1)x^3}{L^3\ln^2(aL)}, &|x|\leq\frac{L}{2}\\
\frac{\pi}{2} +\frac{a}{\ln(2a|x-L|)}, &x>\frac{L}{2},
\end{cases}
\ee
which, as for the $KA$, should only be used in the region of interaction. By using Eq.~\eqref{forcegen}, we get the force
\be\label{forcegluekk}
F^{KK} = -\frac{9a^2}{2L^2\ln^2(aL)}.
\ee
Interestingly, this expression only differs from \eqref{fkknaive} with respect to the coefficients; both of them appear in the bottom panel of Fig~\ref{figforcenew}. By adjusting the coefficients of the above expression, taking $F^{KK}=-60/(L^2\ln^4(20L))$, one can fit the curve obtained by accelerating kinks.

We remark that there are other ways to construct the field profiles using the gluing method. For instance, one can consider the same expressions in Eq.~\eqref{phikagluing} for $|x|>L/2$ and $\phi^{KA} (x)= \pi/2 +\alpha +\beta x^2 + \gamma x^4$ for $|x|\leq L/2$, with $\alpha=-a(4\ln^2(aL)-2\ln(aL)+1)/(4\ln^3(aL))$, $\beta=-2a(\ln(aL)-1)/(L^2\ln^3(aL))$ and $\gamma=-4a/(L^4\ln^3(aL))$ to keep the first and second derivatives continuous. The presence of the $x^4$ term which comes from the Taylor expansion around the origin does not modify the force \eqref{forceglueka}. However, by considering extra terms for the $KK$ pair \eqref{phikkgluing}, i.e., $\phi^{KK}=\pi/2+\mu x +\lambda x^3 +\sigma x^5$, with $\mu=-a(15\ln^2(aL)-6\ln(aL)+2)/(4L\ln^3(aL))$, $\lambda=-2a(5\ln^2(aL)-4\ln(aL)+2)/(L^3\ln^3(aL))$ and $\sigma=4a(3\ln^2(aL)-2\ln(aL)+2)/(L^5\ln^3(aL))$, for $|x|\leq L/2$, the force is not \eqref{forcegluekk}. Instead, we get $F^{KK} = -225a^2/(32L^2\ln^2(aL))$. To verify which approximation works better, one must numerically solve the time-dependent equation of motion \eqref{eom}. This is a deep question which we believe that has to be dealt in future works.

\section{Two-field model}\label{sec3}
We now work with a two-field model which allows for the presence of super long-range structures. We denote the second field by $\chi$ and consider the action
\be\label{action2}
S = \int dx\, dt\left(\frac12f(\chi)\p_\mu\phi\p^\mu\phi +\frac12\p_\mu\chi\p^\mu\chi -V(\phi,\chi)\right).
\ee
In the above expression, $V(\phi,\chi)$ is the potential and $f(\chi)$ is a non-negative function. This model was previously considered in Ref.~\cite{constr4} for some specific functions which allows for modifying the tail of kinks. There, it was shown that one may obtain compact or long-range configurations. Here, we are interested in obtaining super long-range solutions. Since the general equations can be found in Refs.~\cite{constr1,constr3,constr4}, we go straight to the first-order framework which emerges from the BPS bound. To do so, we use the energy density, which is given by
\be\label{rho2}
 \rho=\frac{f(\chi)}{2}\left(\frac{d\phi}{dx}\right)^2 +\frac12 \left(\frac{d\chi}{dx}\right)^2 + V(\phi,\chi).
\ee
If the potential has the form
\be\label{vphichiw}
V(\phi,\chi)=\frac12\frac{1}{f(\chi)}\,\left(\frac{\p W}{\p\phi}\right)^2 + \frac12\left(\frac{\p W}{\p\chi}\right)^2,
\ee
with the auxiliary function $W(\phi,\chi)$ being separable in $\phi$ and $\chi$, in the form $W(\phi,\chi)=G(\phi)+H(\chi)$, one obtains the first-order equations
\be\label{fowxi}
\frac{d\phi}{d\xi}=\pm \frac{d G}{d\phi},\qquad
			\frac{d\chi}{dx}=\pm \frac{d H}{d\chi}.
\ee
The equations with upper/lower sign represent the increasing/decreasing solutions; they are related by the change $x\to-x$, so we only use the equations with the positive sign, for simplicity. In the first-order equation for $\phi$, $\xi$ represents the geometric coordinate that feeds the field $\phi$, given by
\be\label{xicoord}
\frac{d\xi}{dx}= \frac{1}{f(\chi(x))}.
\ee
In the previous works \cite{constr1,constr3,constr4} it was shown that $\xi$ may modify the kink significantly, changing its internal structure and/or its tail. We remark that, even though we are using a two-field model which is invariant under spatial translations, one may obtain a similar first-order equation by considering the presence of impurities, in which the aforementioned invariance is absent; see Ref.~\cite{slawinska}. In this situation, the coordinate $\xi$ can be related to the impurity. Solutions of \eqref{fowxi} minimize the energy of the system, $E = \left|W(v_+,w_+)-W(v_-,w_-)\right|$, where $v_\pm = \phi(\pm\infty)$ and $w_\pm=\chi(\pm\infty)$. 
We work with
\be\label{W}
 W(\phi,\chi) = G(\phi) + \alpha\arctan(\sinh(\chi)),
\ee
where $\alpha$ is a real parameter which must be strictly positive. Notice that the above function is separable in $\phi$ and $\chi$, so the first-order equation for $\chi$ is independent; it reads $d\chi/dx=\sech(\chi$). By solving it with $\chi(0)=0$, we get
\be\label{solvacuumless}
\chi(x) = \arcsinh(\alpha x).
\ee
This function must be used in the left equation of \eqref{fowxi} to calculate $\phi$, which is governed by
\be\label{fophi}
\frac{d\phi}{d\xi}= \frac{dG}{d\phi},\quad\text{where}\quad \xi = \int\frac{dx}{f(\arcsinh(\alpha x))}.
\ee
Therefore, both $f$ and $G$ must be specified to find the solutions. The fact that $\phi$ and $\chi$ do not mix in the auxiliary function \eqref{W} allows for writing the energy density in Eq.~\eqref{rho2} as the sum of two contributions, in the form
$\rho = \rho_\phi + \rho_\chi$, where
\be\label{rhophichi}
\rho_\phi = f(\chi)\left(\frac{d\phi}{dx}\right)^2,\quad \rho_\chi = \left(\frac{d\chi}{dx}\right)^2 = \frac{\alpha^2}{1+\alpha^2x^2},
\ee
in which we have used the solution \eqref{solvacuumless} to obtain the expression in the right equation. By integrating $\rho_\chi$ in all the space, we get the energy $E_\chi=\alpha\pi$. 

Next, we use this formalism to transform solutions into super long-range ones. First, we go from the long-range solution \eqref{solcos4} that arises in the one-field model described by the potential \eqref{vcos4}, to the super long-range solution \eqref{solsuper}. Then, we go from the sine-Gordon solution \eqref{solsg} to a different super long-range solution.

\subsection{From long- to super long-range structures}
Notice that the first-order equation in \eqref{fophi} has the same form of \eqref{fo} with the change $x\to\xi$. This allows us to relate $G_\phi$ with the potential of the one-field model \eqref{action}. Using this approach with the potential in Eq.~\eqref{vcos4}, we consider that $\phi$ in our two-field model is given by
\be
G(\phi) = \frac{1}{2}\left(\phi+\frac12\sin(2\phi)\right).
\ee
In this case, the solution of \eqref{fophi} is similar to \eqref{solcos4}, in the form
\be\label{solphixi1}
\phi(x) = \arctan(\xi(x)).
\ee
Notice, however, that the argument of the function depends on $\xi$, which must be calculated from the integral in Eq.~\eqref{fophi} for the function $f$ under investigation. In the case $\xi(x)=x$, which is equivalent to $f=1$, the solution \eqref{solcos4} is fully recovered. We then take
\be
f(\chi) = \frac{1+\beta}{1+\beta\,\sech(\chi)},
\ee
where $\beta$ is a non-negative real parameter. For $\beta=0$ ($f=1$), $\phi$ and $\chi$ decouples in the Lagrangian density, so the solution \eqref{solphixi1} reduces to \eqref{solcos4}. In this sense, $\beta$ controls how much the field $\chi$ modifies $\phi$ via the function given above. In the limit $\beta\to\infty$, we obtain $f(\chi)=\sinh(\chi)$. For general $\beta$, we get from Eq.~\eqref{fophi} that
\be\label{xilongtosuper}
\xi(x) = \frac{1}{1+\beta}\left(x+\frac{\beta}{\alpha}\arcsinh(\alpha x)\right).
\ee
For $x$ very large, we get the behavior
\be\label{xiasylongtosuper}
\xi(x) = \frac{x}{1+\beta} +\frac{\beta\ln(2\alpha x)}{\alpha(1+\beta)} + \frac{\beta}{4\alpha^3(1+\beta)x^2}+ \mathcal{O}\left(\frac{1}{x^4}\right)
\ee
and
\be\label{solasylongtosuper}
\begin{aligned}
\phi(x) &= \frac{\pi}{2} -\frac{1+\beta}{x} +\frac{\beta(1+\beta)\ln(2\alpha x)}{\alpha x^2}\\
    &-\frac{1+\beta}{x^3}\left(\frac{\beta^2\ln^2(2\alpha x)}{\alpha^2}-\frac{(1+\beta)^2}{3}\right) +\mathcal{O}\left(\frac{1}{x^4}\right).
\end{aligned}
\ee
Therefore, for $\beta=0$, we get $\xi(x)=x$ and, from Eq.~\eqref{solphixi1}, $\phi(x)$ is exactly as in \eqref{solcos4}. As $\beta$ gets larger and larger, the power-law contribution in the geometric coordinate \eqref{xiasylongtosuper} becomes less and less relevant. However, the above expression for the asymptotic behavior of the solution is not valid for the limit $\beta\to\infty$, which must be dealt with some care. The general behavior of $\phi(x)$ for $\beta\to\infty$ has the same form in Eq.~\eqref{solsuper} with the change $a\to\alpha$ and the asymptotic behavior of $\xi$ is $\xi(x) = (1/\alpha)\ln\big(2\alpha x\big) +\mathcal{O}\big(x^{-2}\big)$ and the tail of the solution has the same form in Eq.~\eqref{solsuperasy}. Therefore, the parameter $\beta$ continuously modifies the tail of the solution until it attains a super long-range profile.
%%%%%%%%%%%%%%%%%%%%%%%%%%%%%%
\begin{figure}[t!]
\centering
\includegraphics[width=0.8\linewidth]{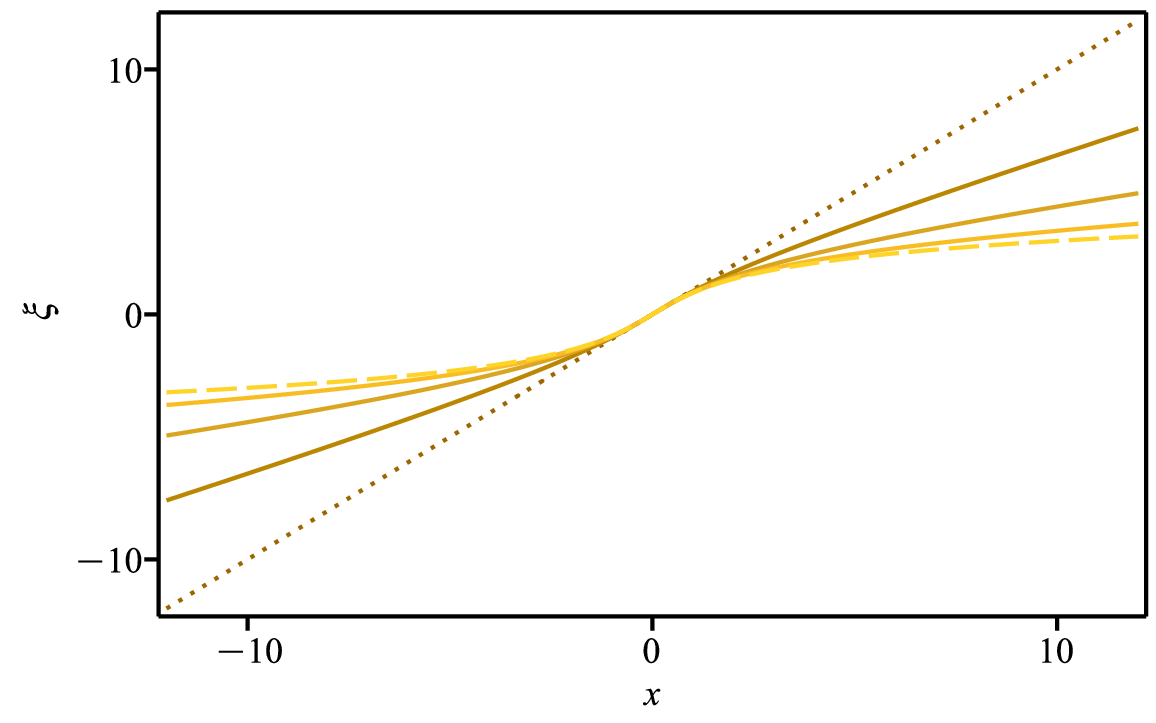}
\includegraphics[width=0.8\linewidth]{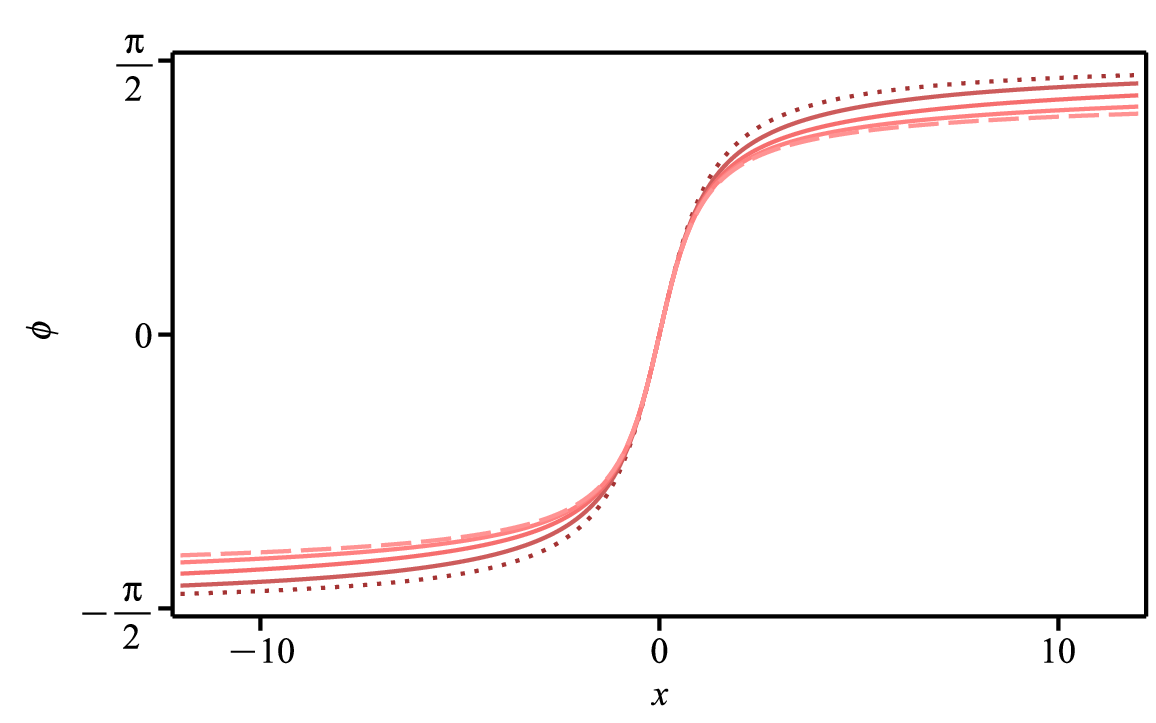}
\caption{The geometric coordinate $\xi(x)$ in \eqref{xilongtosuper} (top, yellow colors) and the associated solution $\phi(x)$ in \eqref{solphixi1} (bottom, red colors) with $\alpha=1$, for $\beta=0,1,4,16$ and $\beta\to\infty$. The dotted lines represent the case supporting the long-range structure ($\beta=0$) and the dashed ones stand for the super long-range configuration. In each panel, the colors get lighter as $\beta$ increases.}
\label{fig7}
\end{figure}
%%%%%%%%%%%%%%%%%%%%%%%%%%%%%%
In Fig.~\ref{fig7}, we display the transition from long- to super long-range solutions, depicting the solution \eqref{solphixi1} for several values of $\beta$.

\begin{figure}[t!]
\centering
\includegraphics[width=0.8\linewidth]{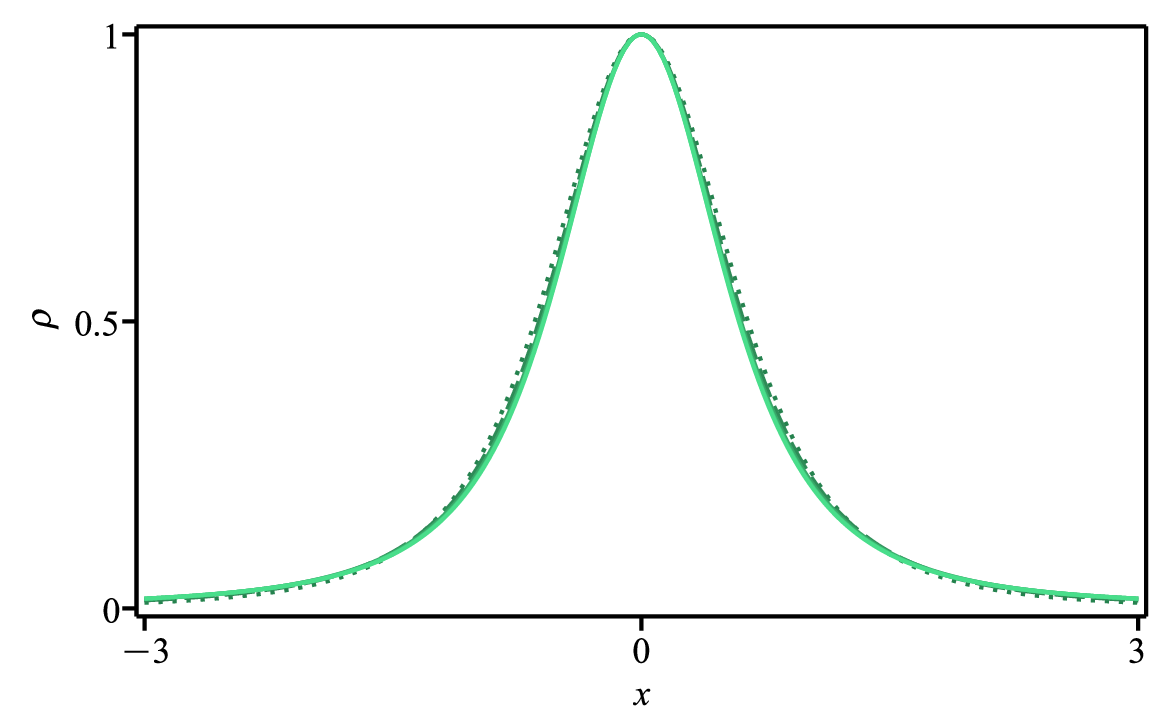}
\includegraphics[width=0.5\linewidth]{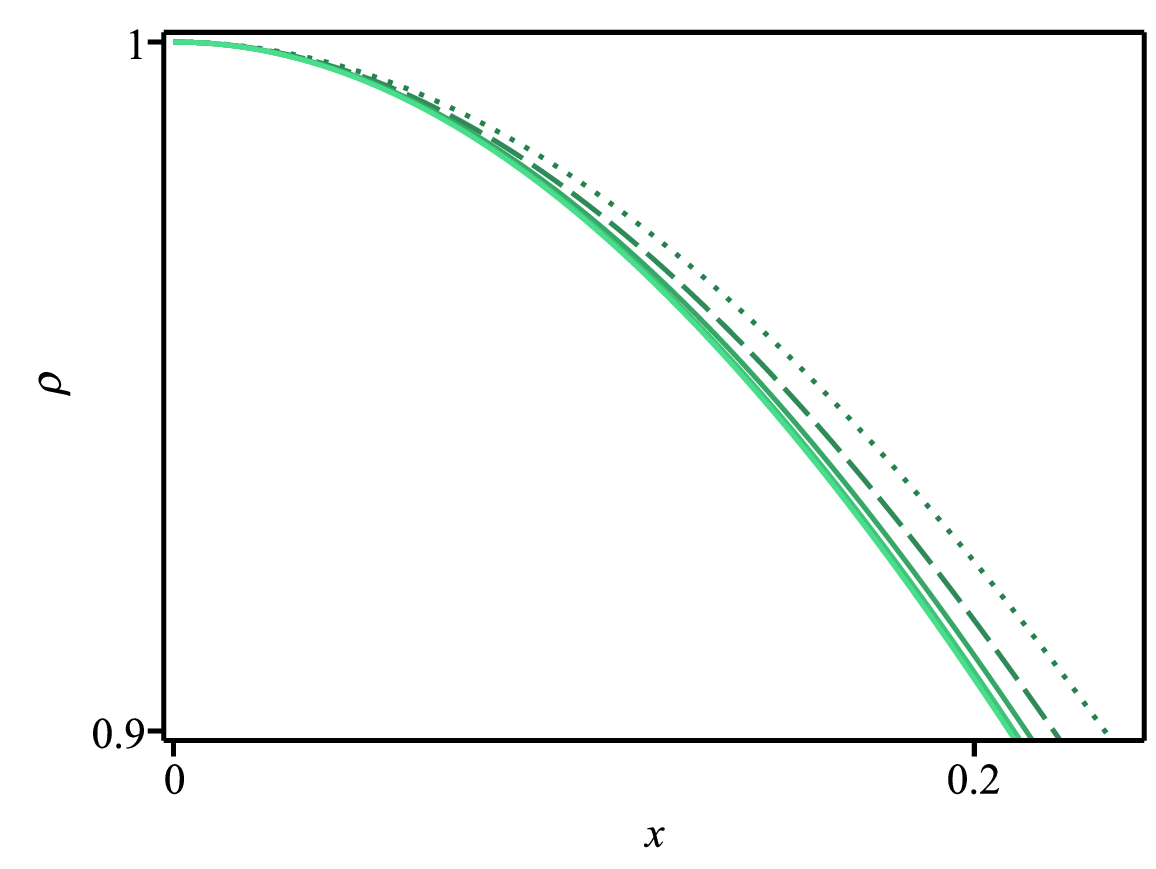}\includegraphics[width=0.5\linewidth]{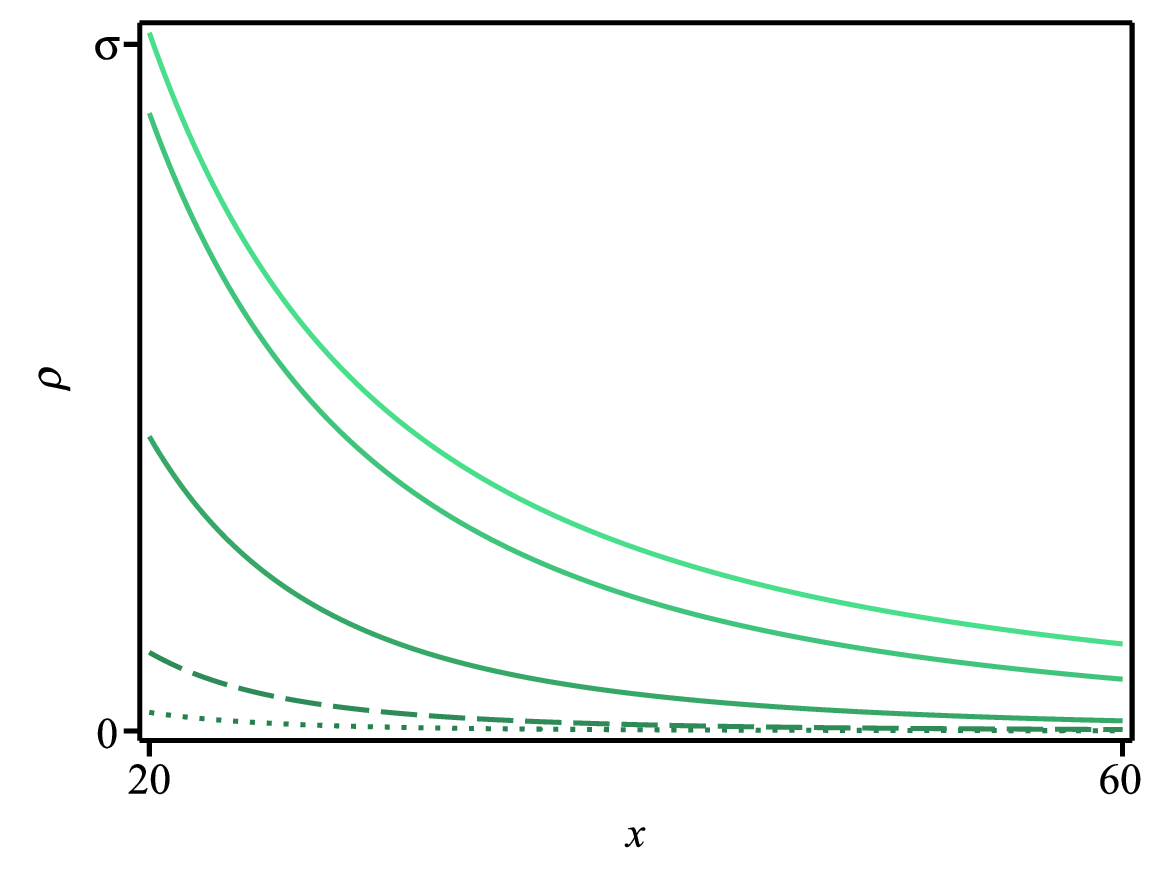}
\caption{The energy density \eqref{rhophilongtosuper} with $\alpha=1$, for $\beta=0,1,4,16$ and $\beta\to\infty$. In the top panel, we display the general behavior. In the bottom-left panel, we display $\rho(x)$ near the origin and, in the bottom-right panel, we depict the asymptotic behavior with the scale of the vertical axis defined by $\sigma=2.3\times 10^{-4}$. The line styles follow Fig.~\ref{fig7}. The green colors get lighter as $\beta$ increases.}
\label{fig8}
\end{figure}

The contribution associated to the field $\phi$ in the energy density can be obtained from \eqref{rhophichi}, which leads us to
\be\label{rhophilongtosuper}
\rho_\phi = \frac{\alpha^4(1+\beta)^3\left(\beta+\sqrt{1+\alpha^2x^2}\right)}{\sqrt{1+\alpha^2x^2}\!\left(\!\alpha^2{(1+\beta)}^2+{(\alpha x +\beta\,\arcsinh(\alpha x))}^2\!\right)^2}.
\ee
Near the origin, we have $\rho_\phi(x\approx0)\approx1-(4+\alpha^2\beta/(1+\beta))x^2/2$.
The asymptotic behavior is
\be
\rho_\phi(x) = \frac{(1+\beta)^3}{x^4} -\frac{4\beta(1+\beta)^3\ln(2\alpha x)}{\alpha x^5} +\mathcal{O}\left(\frac{1}{x^6}\right).
\ee
Similarly to the solution, the limit $\beta\to\infty$ is special; we have
\be
\begin{aligned}
\rho_\phi(x) &= \frac{\alpha^3}{x(\alpha^2+\ln^2(2\alpha x))} -\frac{\alpha}{x^3}\Bigg(\frac{1}{2\big(\alpha^2+\ln^2(2\alpha x)^2\big)}\\
    &+\frac{\ln(2\alpha x)}{\big(\alpha^2+\ln^2(2\alpha x)^3\big)}\Bigg) +\mathcal{O}\left(\frac{1}{x^5}\right).
\end{aligned}
\ee
The energy density \eqref{rhophilongtosuper} is shown in Fig.~\ref{fig8} for some values of $\beta$, showing the transition undergone by the structure. We see that, $\beta$ controls how fast the geometric coordinate go to infinity, and this affects the tail of the solution, which engenders logarithmic tails for $\beta\to\infty$. 

By integrating the energy density \eqref{rhophilongtosuper}, we get $E_\phi=\pi/2$, so the total energy of the two-field model is $E= E_\phi+E_\chi=(1+2\alpha)\pi/2$. We remark that the energy does not depend on $\beta$, remaining the same for a given $\alpha$ while $\beta$ continuously deforms the long-range solutions ($\beta=0$) into the super long-range structures ($\beta\to\infty$). Furthermore, despite the solution in the limit $\beta\to\infty$ be the very same of \eqref{solsuper}, its associated energy density $\rho_\phi$ and energy $E_\phi$ are different from the corresponding ones in the one-field model; this is due to the presence of the function $f(\chi(x))=\sqrt{1+\alpha^2x^2}$ in the system. 
%%%%%%%%%%%%%%%%%%%%%%%%%%%%%%
%%%%%%%%%%%%%%%%%%%%%%%%%%%%%%

\subsection{From short- to super long-range structures}
\begin{figure}[t!]
\centering
\includegraphics[width=0.8\linewidth]{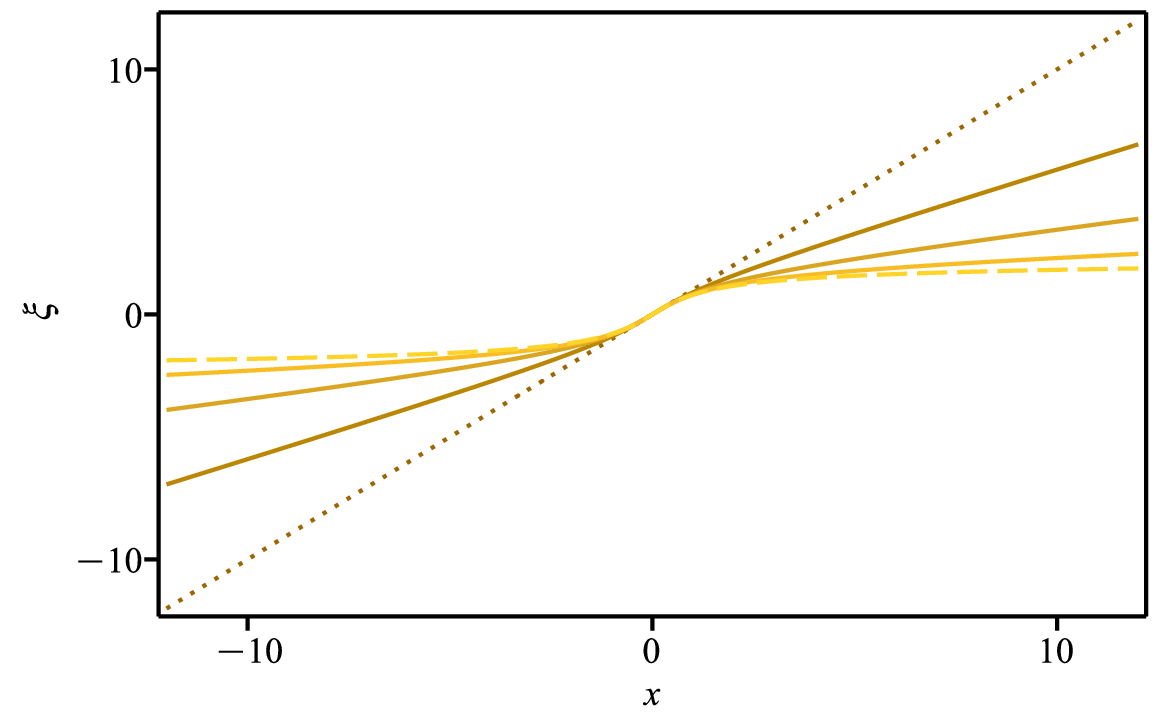}
\includegraphics[width=0.8\linewidth]{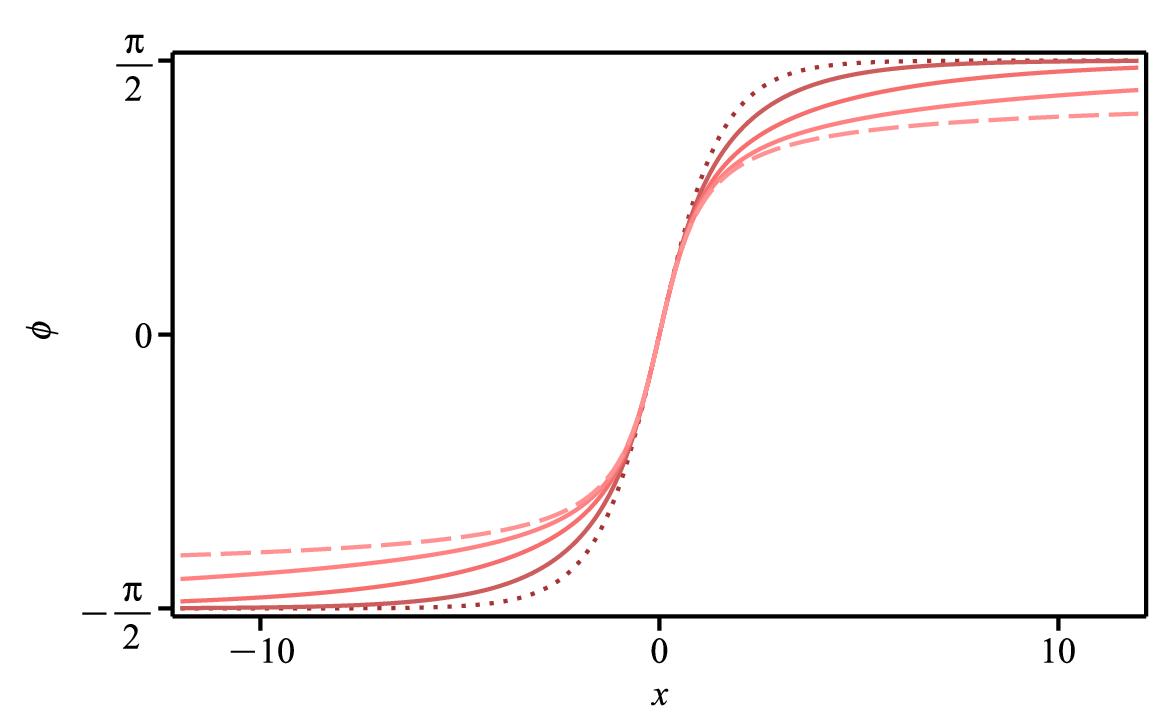}
\caption{The geometric coordinate $\xi(x)$ in \eqref{xishorttosuper} (top, yellow colors) and the associated solution $\phi(x)$ (bottom, red colors) in \eqref{solphixi2} with $\alpha=1$, for $\beta=0,1,4,16$ and $\beta\to\infty$. The dotted lines represent the case supporting the short-range structure ($\beta=0$) and the dashed ones stand for the super long-range configuration. In each panel, the colors get lighter as $\beta$ increases.}
\label{fig9}
\end{figure}

\begin{figure}[t!]
\centering
\includegraphics[width=0.8\linewidth]{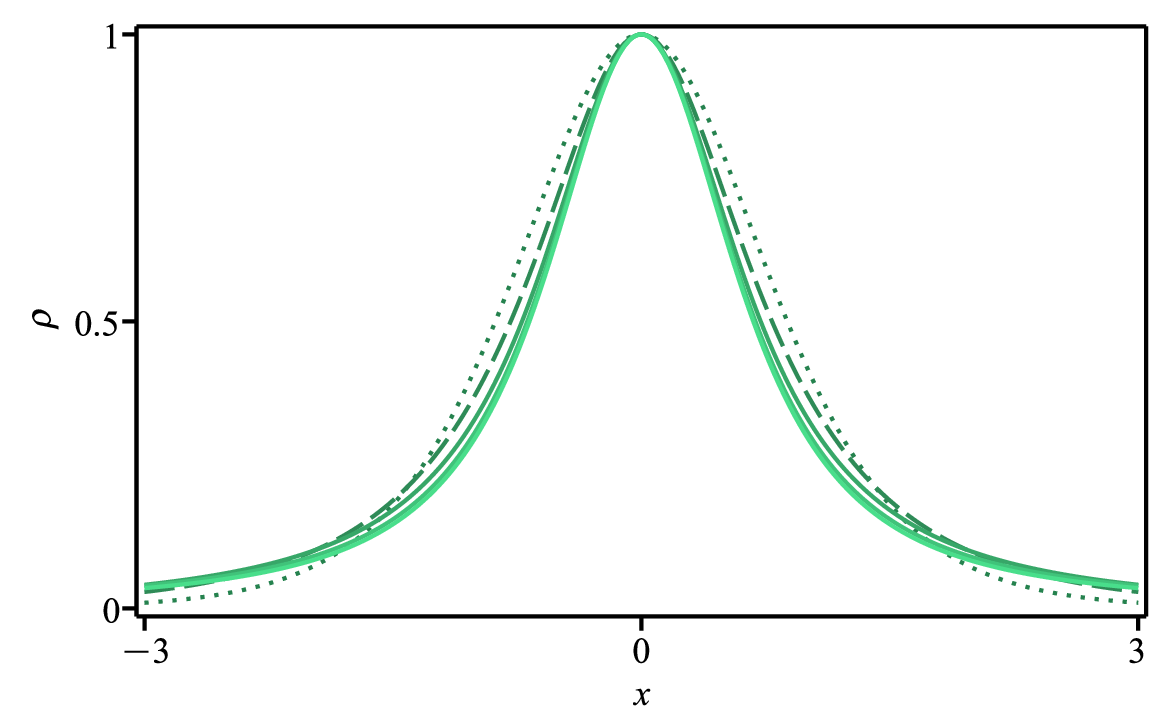}
\includegraphics[width=0.5\linewidth]{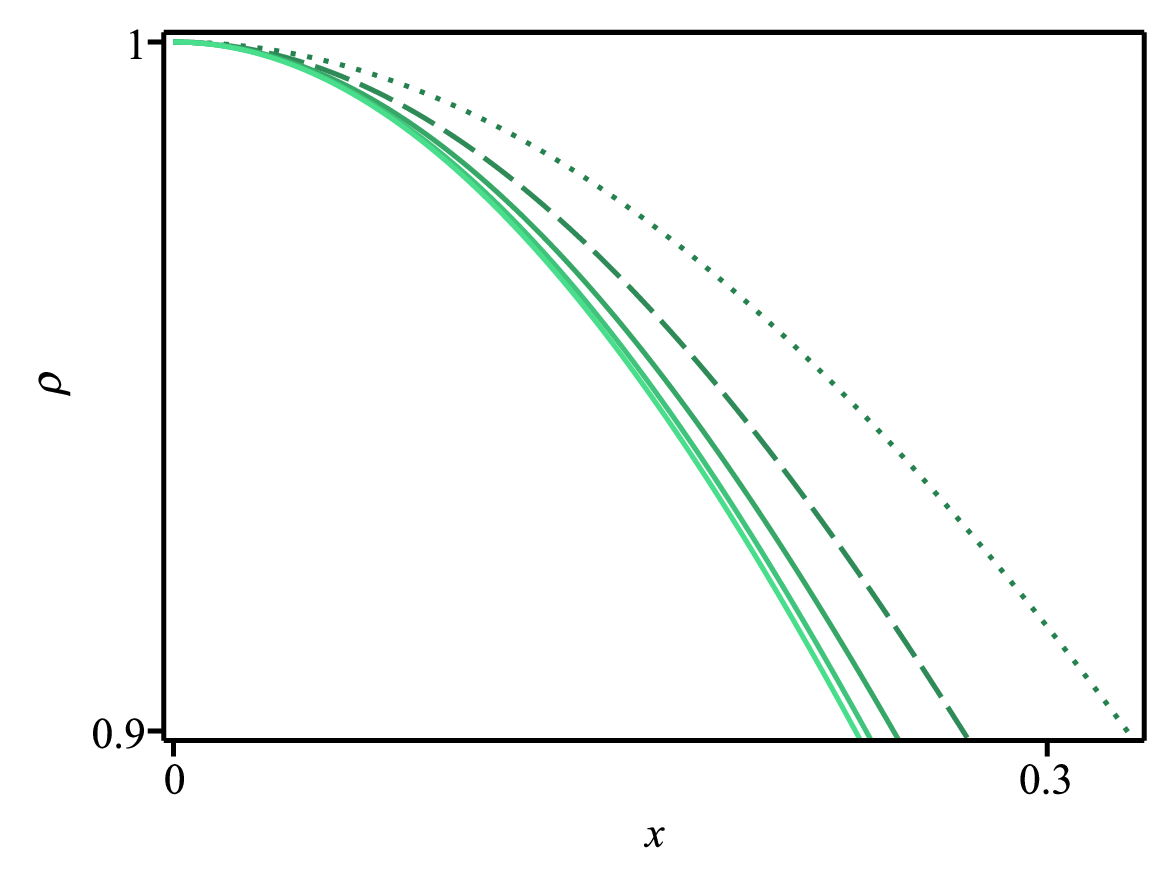}\includegraphics[width=0.5\linewidth]{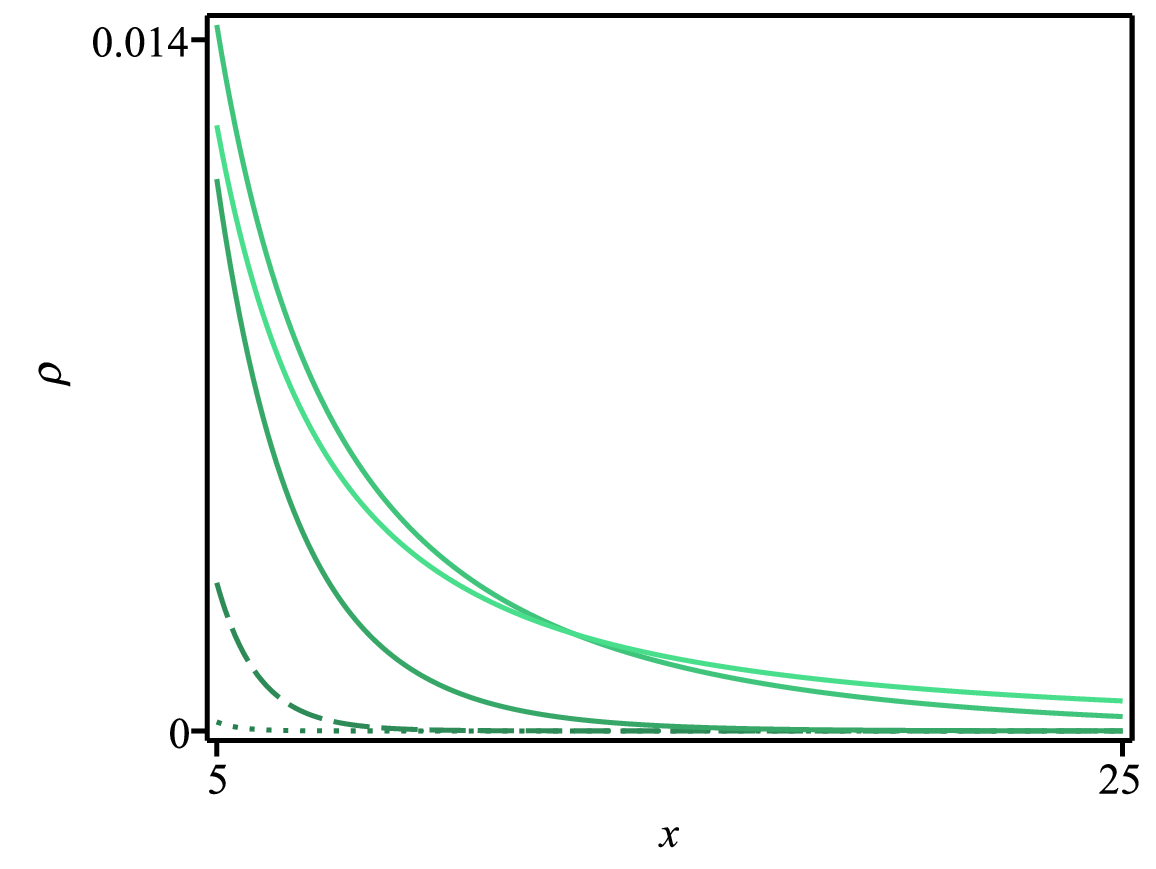}
\caption{The energy density \eqref{rhophishorttosuper} with $\alpha=1$, for $\beta=0,1,4,16$ and $\beta\to\infty$. In the top panel, we display the general behavior. In the bottom-left panel, we display $\rho(x)$ near the origin and, in the bottom-right panel, we depict the asymptotic behavior. The line styles follow Fig.~\ref{fig9} and the green colors get lighter as $\beta$ increases.}
\label{fig10}
\end{figure}

\begin{figure}[t!]
\centering
\includegraphics[width=0.5\linewidth]{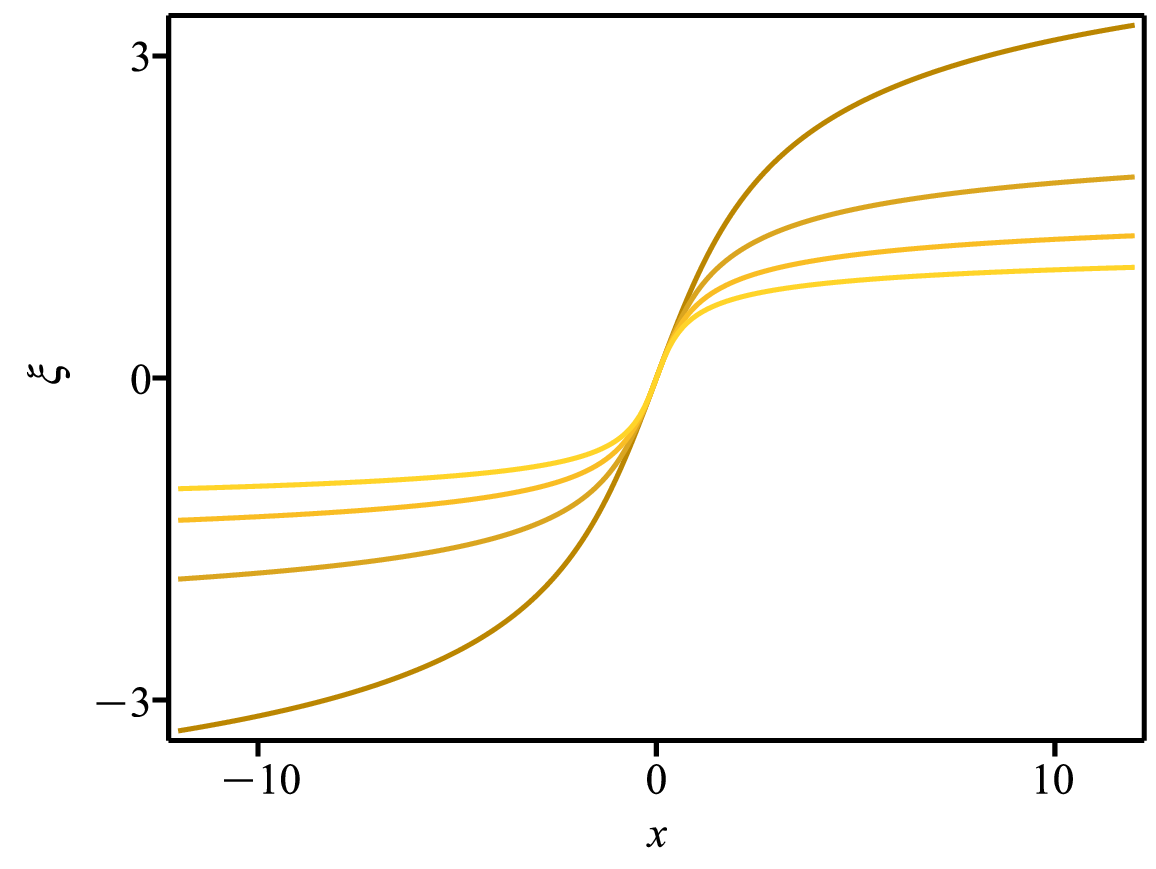}\includegraphics[width=0.5\linewidth]{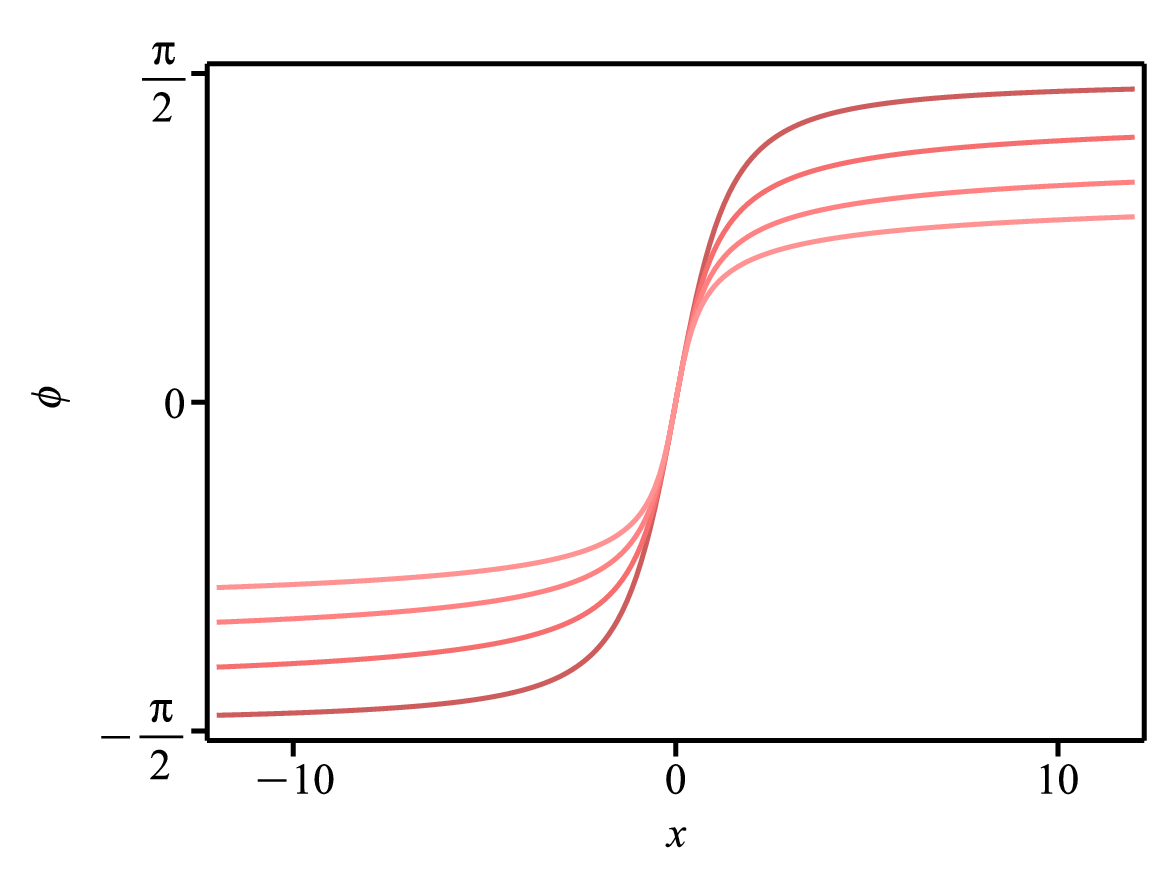}
\includegraphics[width=0.5\linewidth]{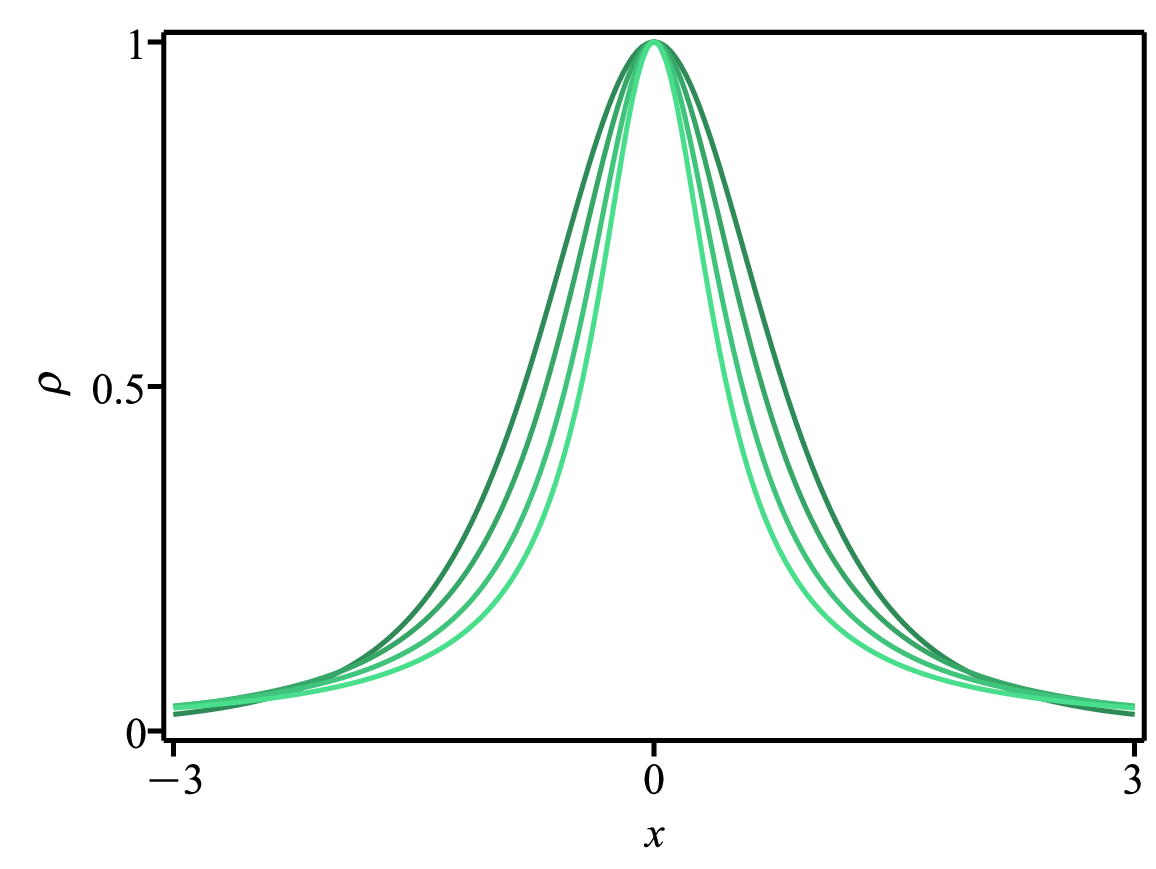}\includegraphics[width=0.5\linewidth]{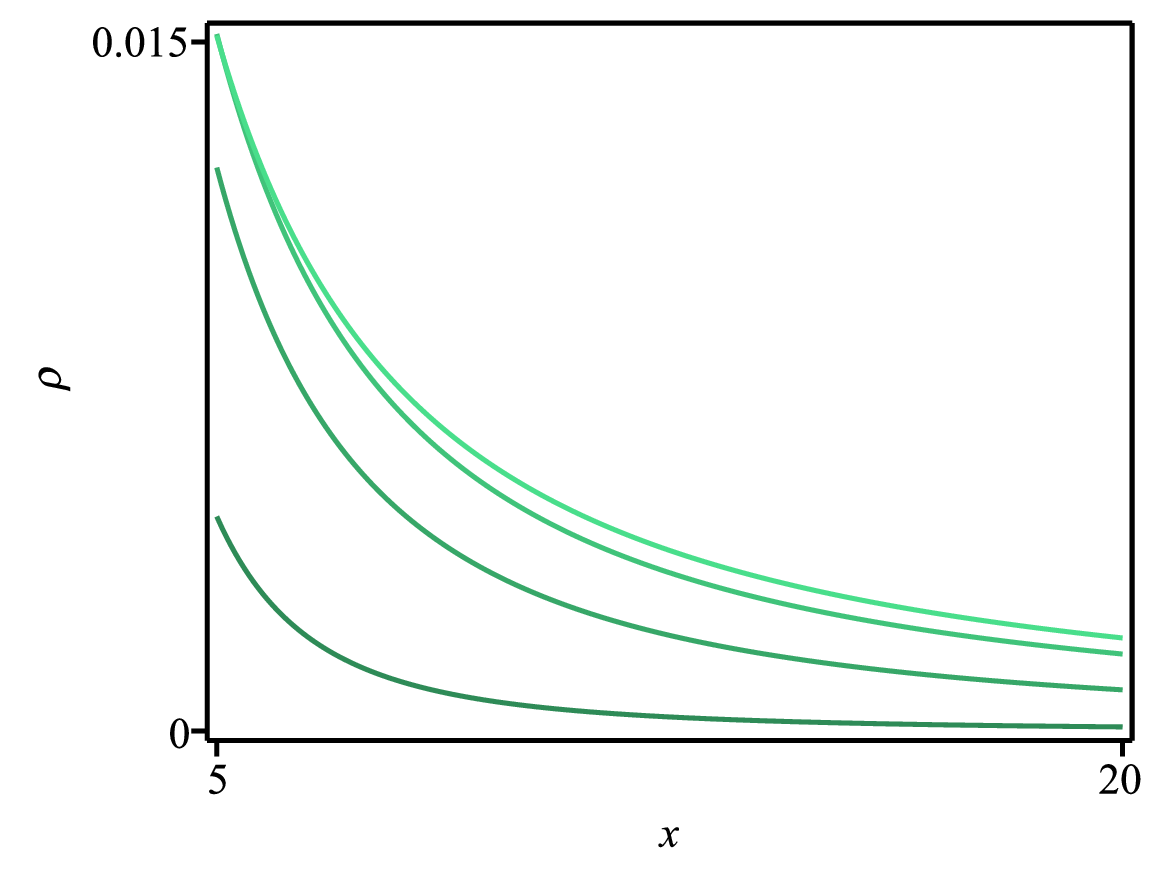}
\caption{The geometric coordinate $\xi(x)$ in \eqref{xishorttosuper} (top left, yellow colors), the associated solution $\phi(x)$ in \eqref{solphixi2} (top right, red colors) and the energy density \eqref{rhophishorttosuper} (bottom, green colors) with $\beta\to\infty$, for $\alpha=1/2,1,3/2$ and $2$. In the bottom-left panel, we show the general behavior of the energy density and, in the bottom-right panel, we display its tail. In each panel, the colors gets lighter with increasing $\alpha$.}
\label{fig11}
\end{figure}

Let us apply our mechanism to smoothly deform the exponentially-tailed sine-Gordon solutions, which we call short range, into super long-range structures. The sine-Gordon model is described by
\be
G(\phi) = \sin(\phi),
\ee
so the solution of Eq.~\eqref{fophi} is similar to \eqref{solsg} with a different argument:
\be\label{solphixi2}
\phi(x) = \arcsin(\tanh(\xi(x))).
\ee
Again, to obtain the geometric coordinate $\xi(x)$, we must solve the integral in Eq.~\eqref{fophi}. The process depends on the function $f(\chi)$. We have found that the transition of our interest is achieved with the function
\be
f(\chi) = \frac{(1+\beta)\sqrt{1+\chi^2}}{\sqrt{1+\chi^2}+\beta\,\sech(\chi)},
\ee
where $\beta\geq0$. For $\beta=0$ ($f=1$), the fields are not coupled. For general $\beta$, we can obtain an analytical expression for the geometric coordinate:
\be\label{xishorttosuper}
\xi(x) = \frac{1}{1+\beta}\left(x+\frac{\beta}{\alpha}\arcsinh(\arcsinh(\alpha x))\right).
\ee
In the case $\beta=0$, the above expression simplifies to $\xi(x)=x$ and \eqref{solphixi2} becomes the usual sine-Gordon solution \eqref{solsg}. On the other hand, for $\beta\to\infty$, we can write $\xi(x)=\arcsinh(\arcsinh(\alpha x))/\alpha$. For general $\beta$, the asymptotic behavior is
\be
\begin{aligned}
\xi(x) &= \frac{x}{1+\beta} +\frac{\beta\ln(2\ln(2\alpha x))}{\alpha(1+\beta)}\\
    &+\frac{\beta}{4\alpha^3(1+\beta)x^2\ln(2\alpha x)} +\mathcal{O}\left(\frac{1}{x^4}\right).
\end{aligned}
\ee
Notice that $\beta$ controls the balance between the power-law and the double-logarithmic terms. The tail of the solution \eqref{solphixi2} is governed by
\be
\phi(x) \approx \frac{\pi}{2} -\frac{2e^{-x/(1+\beta)}}{\ln^{\beta/(\alpha(1+\beta))}\big(4\alpha^2x^2\big)}.
\ee
Therefore, the general behavior is interesting, as it mixes exponential and logarithmic terms. As $\beta$ gets larger and larger, the exponential contribution becomes more and more suppressed. In the limit $\beta\to\infty$, we get a pure-logarithmic tail
\be
\phi(x) \approx \frac{\pi}{2} -\frac{2}{\ln^{1/\alpha}\big(4\alpha^2x^2\big)}.
\ee
It is worth commenting, however, that we have the exponent $1/\alpha$ in the logarithm, which may increase the extension of the super long-range solution. This feature can be seen in Fig.~\ref{fig9}, in which we have plotted the geometric coordinate \eqref{xishorttosuper} and the solution \eqref{solphixi2}. The associated energy density from \eqref{rhophichi} is
\be\label{rhophishorttosuper}
\begin{aligned}
\rho_\phi &= \frac{\beta+\sqrt{1+\alpha^2x^2}\sqrt{1+\arcsinh^2(\alpha x)}}{(1+\beta)\sqrt{1+\alpha^2x^2}\sqrt{1+\arcsinh^2(\alpha x)}}\\
    &\times\sech^2\left(\frac{x}{1+\beta}+\frac{\beta\,\arcsinh(\arcsinh(\alpha x))}{\alpha(1+\beta)}\right).
\end{aligned}
\ee
Near the origin, we get $\rho_\phi(x\approx0)\approx1-(1+\alpha^2-\alpha^2/(1+\beta))x^2$. By integrating the above expression, we obtain $E_\phi=2$. Therefore, the total energy is $E=E_\phi+E_\chi=2+\alpha\pi$. In Fig.~\ref{fig10}, we display this energy density for $\alpha=1$ and several values of $\beta$. Far away from the origin, we get the asymptotic behavior
\be
\rho_\phi(x) \approx \frac{4e^{-2x/(1+\beta)}}{(1+\beta)\ln^{2\beta/(\alpha(1+\beta))}\big(4\alpha^2x^2\big)}\left(1+\frac{\beta}{\alpha x\ln(2\alpha x)}\right).
\ee
In the limit $\beta\to\infty$, it becomes
\be
\rho_\phi(x) \approx \frac{8}{\alpha x\ln^{(\alpha+2)/\alpha}(4\alpha^2x^2)}.
\ee
Similarly to the solution, the parameter $\alpha$ appears in the exponent of the logarithm. This means that $\alpha$, which comes from the field $\chi$, controls how far the tail of the super long-range structure extends. To illustrate this feature, we consider $\beta\to\infty$ and display the solution \eqref{solphixi2} with the geometric coordinate \eqref{xishorttosuper} and energy density \eqref{rhophishorttosuper} for several values of $\alpha$ in Fig.~\ref{fig11}. We see that, as $\alpha$ gets larger and larger, the tail of the solution and energy density extends farther and farther.
%%%%%%%%%%%%%%%%%%%%%%%%%%%%%%
%%%%%%%%%%%%%%%%%%%%%%%%%%%%%%
%%%%%%%%%%%%%%%%%%%%%%%%%%%%%%
%%%%%%%%%%%%%%%%%%%%%%%%%%%%%%

%%%%%%%%%%%%%%%%%%%%%%%%%%%%%%
%%%%%%%%%%%%%%%%%%%%%%%%%%%%%%

\section{Outlook}\label{sec4}
In this manuscript, we have investigated one- and two-field models which support topological structures with logarithmic falloff, which we call super long range. First, we have considered the model described by the action \eqref{action}. We have studied the classical mass associated to the minima of the potential to show that finite non-null mass leads to exponential tail, illustrated by the sine-Gordon potential \eqref{vsg1}. The long-range profile, described by power-law tails, appears in the case of null classical mass; this is seen in the potential \eqref{vcos4}. However, we went deeper and investigated what happens as higher and higher orders of the derivatives of the potential vanish at the minima. To investigate this, we have consider the sine-Gordon-like potential \eqref{vcosn}. We have concluded that, the higher the order of derivatives vanishing at the minima, the farther the tails extend, with the exponent controlling the power-law falloff depending on this feature.

Even though the potential \eqref{vcosn} allows us to control the orders of vanishing derivatives at the minima through the parameter $n$, one cannot determine the behavior in the limit $n\to\infty$. To investigate this issue, we have introduced the potential \eqref{vsuper}, which engenders all orders of the derivatives null at the minima. We have obtained an exact solution and showed that its asymptotic behavior is described by logarithmic falloff. In the energy density, the tails possess terms which mix logarithmic and power-law demeanor; even though it is faster than the usual long-range configurations, it is still slower than the so-called vacuumless structures. We have also investigated the linear stability of the super long-range solutions, which is described a Schr\"odinger-like equation with stability potential of a volcano profile. We have showed that the zero mode does not support nodes, so the model is stable under small fluctuations. The forces between kink-antikink and kink-kink configurations were also investigated using three different methods: we have concluded that they are described by the inverse of products between power-logarithmic and power-law functions of the separation. We have shown that the force \eqref{fkanaive} obtained with the ansatz \eqref{phika} does not agree in functional dependence of the separation nor in coefficients when compared to the ones obtained from accelerating kinks and the gluing technique (see Fig.~\ref{figforcenew} and Eqs.~\eqref{phikagluing} and \eqref{forceglueka}). The latter two seem to agree in functional dependence, only requiring adjustments in the coefficients, at least in the interval depicted in the top panel of Fig.~\ref{figforcenew}. For the $KK$ pair, all the three methods lead to the same functional dependence, being distinguished only by the coefficients; see Eqs.~\eqref{fkknaive} and \eqref{forcegluekk}. This was shown for the interval studied in Fig.~\ref{figforcenew}. We remark that, even though we have found the forces using three methods, all of them involve approximations. Therefore, this issue brings to light an interesting perspective for future research, related to numerical simulation of dynamics using the time-dependent equation of motion.

The second possibility which we have investigated was with the two-field model \eqref{action2}. We have followed the lines of Ref.~\cite{constr4} and used the presence of the second field to modify the behavior of the kink solution. We have taken advantage of the first-order formalism which allows for minimum-energy configurations in which the first-order equation which governs one of the fields can be decoupled. With this method, we have presented how to go from long- to super long-range structures, transforming the power-law into the logarithmic tail due to the action of a parameter in the function $f(\chi)$. Similarly, we have investigated a model that allows for deforming the short- into super long-range structures, which modifies the exponential into the logarithmic tails.

As continuation for the studies in super long-range structures, one may try to generalize our results to get multi-logarithmic tails. For instance, one may investigate the following potential in the one-field model \eqref{action},
\be
V(\phi) = \frac{1}{2}\cos^4(\phi)\,\sech^2(a\tan(\phi))\,\sech^2(\sinh(a\tan(\phi)))
\ee
and $V(v_i)=0$, where $v_i=(i-1/2)\pi$, with $i\in \mathbb{Z}$. In this situation, we get the solution
\be
\phi(x) = \arctan\left(\frac{\arcsinh(\arcsinh(ax))}{a}\right),
\ee
whose asymptotic behavior is given by $\phi(x)\approx \pi/2 - 1/\ln(2\ln(2ax))$. Perhaps, a path to generalize the above result is by using the deformation method \cite{deform,deform2,ganideform1,ganideform2}. In this direction, the study of interkink forces may become even more trickier, as the multi-logarithmic tails fall off slower than the super long-range ones in \eqref{phiasyln}.

Moreover, in this work we have only investigated sine-Gordon-like potentials. Other perspectives include the study of potentials involving polynomial functions, instead of sine-Gordon-like ones, supporting super long-range configurations, both in one- and two-field models. Since solutions engendering logarithmic tails may exhibit forces with range and strength higher than usual, the study of their collisions is also of interest, as it may reveal distinct bounce windows.

\vspace{1cm}
{\noindent\textbf{Data availability statement}: This manuscript has no associated data.}

\acknowledgments{We would like to thank Laudelino Menezes for discussions. We acknowledge financial support from the Brazilian agencies Conselho Nacional de Desenvolvimento Cient\'ifico e Tecnol\'ogico (CNPq), grants Nos. 402830/2023-7 (MAM and RM), 306151/2022-7 (MAM) and 310994/2021-7 (RM), and Paraiba State Research Foundation (FAPESQ-PB) grant No. 2783/2023 (IA).}

%\appendix

%\bibliography{biblio} 

\end{document}